\documentclass[11pt]{article}
\usepackage{geometry}                
\usepackage{graphicx}
\usepackage{amssymb}
\usepackage{amsmath}
\usepackage{epstopdf}
\usepackage{color}
\newtheorem{defi}{Definition}[section]

\newtheorem{prop}[defi]{Proposition}

\newtheorem{Theo}{Theorem}[section]
\newtheorem{Def}[Theo]{Definition}

\newtheorem{Prop}[Theo]{Proposition}

\newcommand{\DEM}{\noindent {\bf Proof.}\ \ }
\newcommand{\QED}{\ \hfill \rule{0.5em}{0.5em} }
\newcommand{\DEF}{\stackrel{\mbox{def}}{=}}

\newcommand{\be}{\begin{equation}}
\newcommand{\en}{\end{equation}}
\def\ad{a^{\dag}}
\def\H{\mathcal{H}}
\def\R{\mathbb{R}}
\def\N{\mathbb{N}}
\def\C{\mathbb{C}}

\def\Z {\mathbb{Z}}
\def\I{\mathbb{I}}
\def\NN{\mathcal{N}}
\def\X{\mathcal{X}}
\def\ZT{\Z_{\tau}}

\def\xa{x^{\ast}}

\newcommand\deq{\stackrel{\mathrm{def}}{=}}

\title{Non-commutative reading of the complex plane through  Delone sequences}

\author{S. Twareque Ali$^{\ast}$,  Lubka Balkova$^{\ast\ast}$,
E.M.F.  Curado$^{\ast\ast\ast}$,  J. P. Gazeau$^{\ast\ast}$, \\
M.A.  Rego-Monteiro$^{\ast\ast\ast}$, Ligia M.C.S.
Rodrigues$^{\ast\ast\ast}$ and K.
Sekimoto$^{\ast\ast\ast\ast}($\footnote{e-mail:
stali@mathstat.concordia.ca, l.balkova@centrum.cz, evaldo@cbpf.br,
gazeau@apc.univ-paris7.fr,  regomont@cbpf.br, ligia@cbpf.br, ken.sekimoto@espci.fr} )\\
\emph{$^{\ast}$ Department of Mathematics and Statistics,}\\
      \emph{Concordia University, Montr\'eal, Qu\'ebec, Canada H3G 1M8}\\
\emph{$^{\ast\ast}$ Laboratoire APC,
Universit\'e Paris  Diderot,}   \\
\emph{10, rue A. Domon  et  L. Duquet
75205 Paris Cedex 13, France}\\
\emph{$^{\ast\ast\ast}$ Centro Brasileiro de Pesquisas Fisicas,}\\
\emph{Rua Xavier Sigaud 150, 22290-180 - Rio de Janeiro, RJ, Brazil}\\
\emph{$^{\ast\ast\ast\ast}$Laboratoire MSC\&ESPCI,
Universit\'e Paris  Diderot,}   \\
\emph{10, rue A. Domon  et  L. Duquet 75205 Paris Cedex 13, France}
}

\date{November 17, 2008} 
\begin{document}


\maketitle

\begin{abstract}

The Berezin-Klauder-Toeplitz (``anti-Wick'') quantization or ``non-commutative reading'' of
the complex plane, viewed as the phase space of a particle moving
on the line, is derived from the resolution of the unity provided by the
standard (or gaussian) coherent states. The construction
properties of these states and their attractive properties are essentially
based on the energy spectrum
of the harmonic oscillator, that is on the natural numbers. This work is
an attempt for following the same path by considering sequences of non-negative
numbers which are not ``too far'' from the natural numbers. In particular,
we  examine the  consequences of such perturbations on the non-commutative
reading of the complex plane in terms of its probabilistic, functional, and
localization aspects.

\end{abstract}

\section{Introduction}

Ever since the initial discovery of coherent states by Shr\"{o}dinger in 1926 and
their rediscovery in the early sixties, independently by Glauber and Sudarshan
(in the context of quantum optics) and by Klauder (following a more thorough
examination of the relationship between classical and quantum mechanics), many
striking features and applications of these particular superpositions of
harmonic oscillator eigenstates have been unveiled and explored. Certainly,
one of the most intriguing and deep properties of these states is the
Bayesian duality \cite{alhelgaz} that they encode between the discrete Poisson
probability
distribution, $n \mapsto e^{-\vert z \vert^2} \vert z \vert^2/n!$, of obtaining
$n$ quantum excitations (``photons'' or ``quanta'') in a measurement through
some  counting device, and the continuous Gamma probability distribution
measure $\vert z\vert^2  \mapsto e^{-\vert z \vert^2} \vert z \vert^2/n!$
on the classical
phase space, with $z$ being a parameter defining the coherent states. For  this latter
distribution, $\vert z\vert^2$ is itself a random variable, denoting the
average number of photons, given that $n$ photons have been  counted. Such a
duality  underlies the construction of all types of  coherent state
families, provided they satisfy a resolution of the unity condition. It turns out
that this condition is
equivalent to setting  up a ``positive operator valued measure''
(POVM) on the phase space. Such a measure, in turn,  leads in a natural
way to a quantization procedure  for the classical phase space, namely the
complex plane, a procedure which we also look upon as being
a ``non-commutative reading''
of the complex plane. (The non-commutativity is understood in the sense that
we associate to each phase space point $z$ the one dimensional
projection operator $\mathrm{P}_z$, projecting onto  to the subspace generated by the
coherent state vector, and  then for $z \neq z', \;\; \mathrm{P}_z \mathrm{P}_{z'}
\neq \mathrm{P}_{z'} \mathrm{P}_z$).
This ``Berezin-Klauder-T\"oplitz'' quantization turns out, in this case, to be
equivalent to the canonical quantization procedure. Clearly, this non-commutative
reading of the
complex plane is intrinsically based on the nonnegative integers (appearing in the
$n!$ term). In this paper
we will  follow a similar path by considering sequences of non-negative numbers
which are not ``too far'' from the natural numbers. The resulting quantizations
will then be looked upon as perturbations of the non-commutative reading given by the
standard coherent states. In particular, we will
examine the  consequences of such perturbations on the non-commutative reading
of the complex plane as they relate to their probabilistic, functional, and
localization properties.

More precisely, let us consider  an infinite, strictly increasing sequence of
nonnegative real numbers
$\{ x_n\}_{n\in \N}$ such that $x_0 = 0$ and there
exist $r,\; L \in \mathbb R$ such that $0 < r \leq x_{n+1} - x_n
\leq L < \infty $ for any $n \in \N$ (``Delone sequence'').
To this sequence of numbers correspond the sequence of ``factorials''
$x_n! = x_1x_2\dotso x_n$ with  $x_0! = 1$, the ``exponential''
$$
\mathcal{N}(t) = \sum_{n = 0}^{+\infty} \frac{t^n}{x_n!}\, ,
$$
the sequence of ``moment'' integrals,
$$
x_n! \mu_n =\int_{0}^{+\infty} \frac{t^n}{\mathcal{N}(t)}\, dt\, .
$$
and the ``renormalized" sequence
$$
\tilde{x}_n := \frac{\mu_n}{\mu_{n-1}}\, x_n\, , \ n\geq 1\, , \quad \tilde{x}_0 = 0\, .
$$

In the present study we consider various examples of such sequences, for
which it is proved, analytically or numerically, that the ratios
$ \dfrac{\mu_n}{\mu_{n-1}}\to 1$.

The ``non-commutative reading'' of the complex plane is understood in the
following sense. Standard coherent states are vectors labelled by
$z \in \C$,  $v_z = e^{-\vert z \vert^2/2}\sum_{n\in \N} z^n/\sqrt{n!}\, e_n$,
in some separable Hilbert space with orthonormal basis $\{e_n\, , \, n\in \N\}$.
One of their most striking features is that they solve the unity:
\begin{equation*}
\I ={\displaystyle\int\nolimits_{\C} } \dfrac{d^2 z}{\pi}\, (v_z\, , \cdot)v_z\,.
\end{equation*}
This property allows us to transform ordinary functions $f(z)$,
 $z = (1/\sqrt{2})( q  + ip)$, into operators
 $f \mapsto A_f = {\displaystyle\int\nolimits_{\C} }
 \dfrac{d^2 z}{\pi}\, f(z) (v_z\, , \cdot)v_z$ (``coherent state quantization'')
 and yields a non-commutative version of the  plane illustrated by
 $[Q,  P] = i \I$, where $ Q \equiv A_q$ and $ P \equiv A_p$. Moreover,
 these states minimize that non-commutativity in the sense that  the uncertainty
 relations (or Heisenberg inequalities) for the quantized coordinates $Q$ and $P$ are saturated:
\begin{equation*}
\Delta Q \, \Delta P = \frac{1}{2}\, , \quad  \Delta Q \deq
\sqrt{(v_z, Q^2 v_z) - (v_z, Q v_z)^2 }\, .
\end{equation*}
Such an equality produces a ``fuzzy'' vision of the plane with elementary
``minimal'' square lattice cells of side length $1/\sqrt{2}$.

Deformed coherent states are constructed using the renormalized sequence
$\tilde{x}_n := \mu_n/\mu_{n-1}\, x_n \, .$
$\tilde{v}_z = \sum_{n=0}^{\infty} 1/\sqrt{\widetilde{\mathcal{N}}(\vert z
\vert^2)}\, z^n/\sqrt{\tilde{x}_n!}\,  e_n\, .$ They also yield a  resolution of the identity
\begin{equation*}
\I = \int_{\C} \frac{\widetilde{\mathcal{N}}(\vert z \vert^2)}{\mathcal{N}
(\vert z \vert^2)}\,\dfrac{d^2 z}{\mu_0\, \pi} \, (\tilde{v}_z\, , \cdot)\tilde{v}_z\, ,
\ \widetilde{\mathcal{N}}(t) = \sum_{n = 0}^{+\infty} \frac{t^n}{\tilde{x}_n!}\, ,
\end{equation*}
and, by following the same procedure as above,  a non-commutative version of
the plane. For sequences which are perturbations of natural integers in a sense
which will be made precise in the paper, the ratio $ \frac{\widetilde{\mathcal{N}}
(\vert z \vert^2)}{\mu_0\, \mathcal{N}(\vert z \vert^2)}$ should be $\approx 1$. This represents a mild
perturbation of the standard case, with uncertainty relations  modified as
\begin{equation*}
\Delta \widetilde{Q} \, \Delta \widetilde{P} = \frac{1}{2}\,
(\tilde{v}_z, (\tilde{x}_{N+1} - \tilde{x}_N)\tilde{v}_z)\, ,
\end{equation*}
where the diagonal operator $\tilde{x}_N$ is defined by
$\tilde{x}_N\, e_n = \tilde{x}_n\, e_n$. If the differences $\tilde{x}_{n+1} - \tilde{x}_n$
are not too far from 1, then we deal with a kind of curved
deformation of the square
lattice of uncertainty cells corresponding to the standard case.

Therefore, the aim of the paper is to explain  how such  sequences of numbers allow
one to implement
a specific coherent state quantization from the complex plane,
different from the canonical quantization associated with the standard
integers. Accordingly, consequences in terms of phase space localization  are examined.

The rest of this article is organized as follows. In Section 2, we define what we
call Delone sequences of nonnegative numbers, more precisely Delone
$\alpha$-perturbations of nonnegative integers, and we give a list of
examples of such sequences.
We revisit in  Section 3 the quantization
of the complex plane by using the ordinary (or Gaussian) coherent states.
We then attempt in Section 4  to follow the same quantization procedure
but this time based on an $\alpha$-perturbation of the natural numbers. The
existence of a POVM is dependent on the associated Stieltjes moment problem
according to whether the latter does  or does not have an explicit  solution.
Even when a solution exists,
it is interesting to adopt an alternative strategy for constructing a POVM which,  in
a certain sense, is a perturbation of the standard coherent states POVM on the
complex plane.
This strategy is based on a ``renormalization'' of the original Delone
$\alpha$-perturbation of the natural numbers. We then formally proceed in
Section 5 to the quantization of the complex plane arising from this renormalized
sequence. We give in Section 6, some insights about the algebraic structure of
the basic operators (ladder operators and the resulting successive commutators)
emanating from this  quantization. Next, we describe the localization aspects of
the quantization of the complex plane by examining the associated orthogonal
polynomials generalizing the Hermite polynomials (Section 7) and the mean values
(or ``lower symbols")  of operators obtained from this quantization  (Section 8).
Section 9 is devoted to a particularly illuminating example of
$\alpha$-perturbation, namely  a situation in which  generalized
hypergeometric functions are involved and where the probabilistic aspects of
our approach
are clearly brought out. In Section 10, we go back to  the algebraic aspects
of the quantization through $\alpha$-perturbations of natural numbers by examining
the associated generalized Heisenberg algebra, a concept developed by some of
us. In Section 11, we examine in a more mathematically oriented way the conditions
under which it is possible to
suitably  Delone renormalize the $\alpha$-perturbation of the natural numbers,
and we present a new result on asymptotic Poisson and Gamma distributions.

\section{Delone sequences}
Suppose that we are given an infinite strictly increasing sequence of
nonnegative real numbers
\begin{equation}
\label{seq}
\mathcal{X} = \left( x_n\right)_{n\in \N}\, , \qquad x_0 = 0\, ,
\end{equation}
obeying the following two constraints :
\begin{itemize}
  \item[(d1)] $\mathcal{X}$ is \emph{uniformly discrete} on the positive real
  line $\R^+\, : $ $\exists\, r>0$ such that $x_{n+1} - x_n \geq r$ for all $n\in N$,
  which means that there exists a minimal distance between two successive elements
  of the sequence,
  \item[(d2)] $\mathcal{X}$ is \emph{relatively dense} on $\R^+\, :$ $\exists \, R>0$
  such that for all $x \in \R^+$ $\exists\,  n \in \N$ such that $\vert x - x_n\vert < R$,
  which means that there exists a maximal distance, say $L$,  between two successive
  elements of the sequence.
\end{itemize}
These conditions imply that  $\lim_{n \to \infty} x_n = \infty$. Actually, they mean that the
sequence $\mathcal{X}$ forms a  \emph{Delone}  set, and we will call $\mathcal{X}$
a (non-negative) \emph{Delone} sequence  in $\R^+$.  We  could choose the scale in such
a way that $x_1 = 1$ or $L=1$, but we leave this as an option.

The rationale behind this set of constraints is that the sequence $\mathcal{X}$ should
thus appear as not very different from the set of natural numbers. It should be noticed
that from this point of view, sequences like
$x_n = n^{\alpha}\, , \, \alpha \neq 1$, or
$x_n = n\, \log{n}$ are not Delone. Also, familiar deformations of integers, like
$q$-deformations or $(p,q)$-deformations,
\begin{equation}
\label{deform}
[n]_{(p,q)} = \frac{p^{-n}-q^n}{p^{-1}-q}\, , \qquad [n]_q = [n]_{(1,q)}\, ,
\end{equation}
are not Delone. In this sense, these expressions  could be viewed as
\emph{singular} deformations of the set $\N$.

To the sequence $\mathcal{X}$  corresponds the sequence of ``factorials''
$x_n! = x_1x_2\dotso x_n$ with  $x_0! \deq 1$, the ``exponential''
\begin{equation}
\label{expxn}
\mathcal{N}(t) = \sum_{n = 0}^{+\infty} \frac{t^n}{x_n!}\, ,
\end{equation}
and the sequence of ``moment'' integrals
\begin{equation}
\label{seqmom}
x_n! \, \mu_n =\int_{0}^{+\infty} \frac{t^n}{\NN(t)}\, dt\, ,
\end{equation}
in which the appearance of the ``corrective'' factors $\mu_n$'s is needed since
there is no reason that the Stieltjes moment problem is solved for a generic pair
$(dt/\mathcal{N}(t), x_n!)$, as  would be the case for $(e^{-t}\, dt, n!)$.

The present study will  more specifically concern  sequences of numbers $\{x_n\}$
behaving asymptotically like natural numbers, $x_n \approx \mathrm{const.}\times n$
at large $n$. For convenience, we will put $\mathrm{const.}= 1$ or equivalently we
will consider the sequence $x_n/\mathrm{const.}$.  We will  impose more on
this ``natural'' behavior, of course. We will require in our main example the
following

{\Def[Delone perturbation of $\N$]
Let $ \N \ni n \mapsto \alpha(n) $ be a bounded function with values in the
interval $(-1,1)$, $\alpha(0) = 0$, and such that its successive jumps
$\alpha(n+1) -\alpha(n)$ have lower bound $r-1$ with $r\in (0,1)$. Then the
Delone sequence
\begin{equation}
\label{delpert}
x_n = n + \alpha(n)\, , \quad n \in \N\, ,
\end{equation}
is called  an $\alpha$-perturbation of the natural numbers.
}

\noindent Note that this $r$ is a Delone lower bound for the sequence
$ \{ x_n\}_{n\in \N}$ in the sense of (d1).

For instance, $\alpha(n)$ could be a constant shift, $\alpha(n) =
\epsilon$ for $n > 0$. In order to fulfill the Delone condition (d1),
we should then have $-1 < \epsilon < 1$ and $0 < r < 1$. Another simple case
that will be considered in this paper is
\begin{equation}
\label{homseq}
0 < n \mapsto \alpha(n) = \frac{an + b}{cn +d}\, ,
\end{equation}
with $-1 < b/d < a/c < 1$ or $-1 < a/c < b/d  < 1$ and $0<r<1$.

The perturbation function could also be periodic, like

\begin{equation}
\label{perseq}
\alpha(n) = \epsilon\, \sin{\omega n}\, ,\quad -1< \epsilon<1\,,
\quad 0< r \leq 1 - 2\vert \epsilon\, \sin{\omega}\vert\, .
\end{equation}

 The function $\alpha(n)$ could even be a random perturbation.

Another interesting example is the set $\Z_{\beta}^+$ of
non-negative beta-integers, i.e. all those positive real numbers
which are polynomial in powers of an irrational real number $\beta >
1$, when they are expanded with respect to the base $\beta$ using the usual greedy
algorithm. When $\beta$ is endowed with specific properties (e.g.
Pisot-Vijayaraghavan (PV) algebraic integers or more generally
belonging to the so-called Parry family), these $\beta$-integers
 form a~quasiperiodic sequence with a finite number of possible
adjacent differences $x_{n+1} - x_n $.  The number of
distinct distances between consecutive $\beta$-integers is finite if
and only if $\beta$ is a~Parry number. It is natural to ask in case
of Parry numbers, whether the corresponding $\beta$-integers behave
similarly to the  integers. Some answers concerning the additive,
multiplicative and diffractive properties of $\beta$-integers already exist (see \cite{GaGo} and references therein). In view of the
``quasi''-periodic distribution of the $\beta$-integers on the real
line, it is pertinent to investigate the extent to which they differ
from ordinary integers in an asymptotic sense, according to the
nature of $\beta$.

  Let us quote here precise results
from~\cite{bagapel} that illustrate the similarity between sets
$\mathbb N$ and $\mathbb Z^{+}_\beta=\{b_n \mid n \in \mathbb N\}$
for $\beta$ being a~Parry number by
mentioning two properties:
\begin{enumerate}
\item The limit $c_{\beta}:=\lim_{n \to \infty}\dfrac{b_n}{n}$
exists and belongs to the extension field $\mathbb Q(\beta)$ on the rational numbers.
\item For $\beta$ being moreover a~PV number such
that its minimal and its Parry polynomial coincide, the modulation
$\alpha(n)=\dfrac{b_n}{c_\beta}-n$ is bounded. Parry polynomials are defined in \cite{bagapel}
\end{enumerate}

\medskip

Let us mention that both of the previous asymptotic characteristics
are known for $\beta$ being a~quadratic Parry unit. We underline
that for quadratic numbers the notions of Parry and PV numbers
coincide. The following proposition providing explicit formulae for
$\beta$-integers for a~quadratic PV unit $\beta$ comes
from~\cite{GaGo}.
\begin{prop}\label{unit_gago} If $\beta$ is a~simple
quadratic PV unit, then
$$
 {\mathbb Z}_{\beta}^+ =  \left\{ b_n = c_{\beta} n  +
\frac{1}{\beta} \, \frac{1 - \beta}{1 + \beta} +
\frac{\beta-1}{\beta}\, \left\{ \frac{n+1}{ 1 + \beta}\right\}, \  n
\in \mathbb N \right\}, \quad \text{where} \ c_{\beta}= \dfrac{1 +
\beta^2}{\beta (1 + \beta)}.
$$

\vspace{0.1cm} If $\beta$ is a~non-simple quadratic PV unit, then
$$
{\mathbb Z}_{\beta}^+ =   \left\{ b_n = c_{\beta} n  +
\frac{1}{\beta} \left\{ \frac{n}{  \beta}\right\}, \  n \in \mathbb
N \right\}, \quad \text{where} \ c_{\beta} = 1 - \dfrac{1}{\beta^2}\, ,
$$
where $\{ x\}$ denotes the fractional part of a nonnegative real number $x$.
\end{prop}

 The simplest example is provided
 by the set $\ZT^+$ of non-negative $\tau$-integers, where
$\beta$ is equal to the golden mean $\tau = (1 + \sqrt{5})/2$. These
$\tau$-integers form a~quasiperiodic sequence with two possible
adjacent differences $x_{n+1} - x_n = 1$ or $1/\tau$. Their exact
expression is given by
$$
 {\mathbb Z}_{\tau}^+ =  \left\{ b_n = c_{\tau}\,  n
- \frac{1}{\tau^4} +
\frac{1}{\tau^2}\, \left\{ \frac{n+1}{\tau^2}\right\}, \  n
\in \mathbb N \right\}, \quad \text{where} \ c_{\tau}= \dfrac{1 +
\tau^2}{\tau^3}  \approx 0.8541\, ,
$$
where $\{ x\}$ designates the fractional part of a~nonnegative real
number $x$. Dividing by $c_{\tau}$ gives the ``normalized'' Delone
sequence $ x_n = b_n/c_{\tau} = n + \alpha_{\tau}(n)\, , $ with
$$\alpha_{\tau}(n) = \frac{\tau}{1 + \tau^2}\, \left\{
\frac{n+1}{\tau^2}\right\}  - \frac{1}{\tau(1+\tau^2)} \, .$$ Since
$\{x\} \in [0,1)$, we deduce the bounds for the modulation:
$$
- \frac{1}{\tau(1+\tau^2)} \approx -0.1708  <\alpha_{\tau}(n) < \frac{1}{(1+\tau^2)}
\approx 0.2764\, .
$$

We will explain below how such  Delone sequences of numbers allow one
to implement a specific non-commutative (or ``quantum'') reading of the complex
plane associated to Poisson-like probability distributions and to what extent
this reading is ``close'' to the ``canonical'' reading  provided by natural
numbers. We will start with the sequence of natural numbers and then  we will
generalize the procedure to sequences $\X$ endowed with suitable properties.

\section{Natural numbers and the ``fuzzy plane''}

Let $\H$ be a separable (complex) Hilbert space with scalar
product $(\cdot\, ,\cdot)$ (antilinear on the left) and orthonormal basis
$e_0,e_1,\dotsc, e_n, \dotsc$,
\begin{equation}
\label{scprod}
(e_n,e_{n'}) = \delta_{nn'}\, .
\end{equation}

To each complex number $z \in \C$ corresponds the following vector in $\H$ :
\begin{equation}
\label{stcs}
v_z = \sum_{n=0}^{\infty} e^{-\frac{\vert z \vert^2}{2}}\frac{z^n}{\sqrt{n!}}\,  e_n\, .
\end{equation}

Such a ``continuous'' set of vectors is called \emph{a family of standard coherent states}
 in the quantum physics
literature \cite{klauskag,csbook}. They enjoy the following properties.
\begin{itemize}
  \item[(i)] $\Vert v_z\Vert = 1$ (\textbf{normalization}).
  \item[(ii)] The map $\C \ni z \mapsto v_z$ is weakly
continuous (\textbf{continuity}).
  \item[(iii)] The map $\N \in n \mapsto \vert (e_n\, ,v_z)\vert^2 =
  e^{-\vert z \vert^2} \dfrac{\vert z \vert^{2n}}{n!}$ is a Poisson
  probability distribution with average number of successes equal
  to $\vert z \vert^2$  (\textbf{discrete probabilistic content}
  \cite{alhelgaz}).
\item[(iv)] The map $\C \ni z \mapsto \vert (e_n\, ,v_z)\vert^2 =
e^{-\vert z \vert^2} \dfrac{\vert z \vert^{2n}}{n!}$ is a Gamma
probability distribution (with respect to the square of the radial
variable) with $n$ as a shape parameter (\textbf{continuous
probabilistic content} \cite{alhelgaz}).
\item[(v)] \begin{equation}
\label{stresunity}
\I ={\displaystyle\int\nolimits_{\C} } \dfrac{d^2 z}{\pi}\,  \mathrm{P}_z\, ,
\end{equation}
where
$\mathrm{P}_z = (v_z\, , \cdot)v_z$ is the orthogonal projector onto the
vector $v_z$ and the integral should be understood in the weak sense
(\textbf{resolution of the unity in $\H$}).
 \end{itemize}
The proof of (\ref{stresunity}) is straightforward and stems from the
orthogonality of the Fourier exponentials and from the integral expression
of the Gamma function  which solves the moment problem for the factorial $n!$ :
\begin{equation}
\label{stresunityproof}
{\displaystyle\int\nolimits_{\C} } \dfrac{d^2 z}{\pi}\,  \mathrm{P}_z =
\sum_{n, n'= 0}^{\infty}  e_n\,  (e_{n'}\, ,\cdot) \, \frac{1}{\sqrt{n!n'!}}\,
\int_{\C}\dfrac{d^2 z}{\pi}\,  e^{-\vert z \vert^2}z^n{\bar z}^{n'} =
\sum_{n= 0}^{\infty}  e_n\,  (e_{n}\, ,\cdot) = \I\, .
\end{equation}

Property (v) is crucial in the sense that it allows one to define:

\begin{enumerate}
  \item a normalized positive operator-valued measure (POVM) on the
  complex plane equipped with its Lebesgue measure $\dfrac{d^2 z}{\pi}$
  and its $\sigma-$algebra $\mathcal{F}$ of Borel sets :
\begin{equation}
\label{stpov}
\mathcal{F} \ni \Delta \, \mapsto \int_{\Delta}\dfrac{d^2 z}{\pi}\,
\mathrm{P}_z \, \in \mathcal{L}(\mathcal{H})^+\, ,
\end{equation}
where $\mathcal{L}(\mathcal{H})^+$ is the cone of positive bounded
operators on $\H$;
  \item a so-called Berezin-Klauder-Toeplitz quantization of the complex
  plane \cite{klau1,ber,gghlrl,klau2}, which means that to a function
  $f(z,\bar{z})$ in the complex plane there corresponds the operator
  $A_f$ in $\H$ defined by
  \begin{equation}
\label{stquant}
f \mapsto A_f = \int_{\C}\dfrac{d^2 z}{\pi}\, f(z,\bar{z}) \mathrm{P}_z =
\sum_{n, n'= 0}^{\infty} \left(A_f\right)_{nn'} e_n\,  (e_{n'}\, ,\cdot) \,,
\end{equation}
with matrix elements
\begin{equation}
\label{matel1}
\left(A_f\right)_{nn'}= \frac{1}{\sqrt{n!n'!}}\, \int_{\C}\dfrac{d^2 z}{\pi}\,
f(z,\bar{z}) e^{-\vert z \vert^2}z^n{\bar z}^{n'}\, ,
\end{equation}
 provided that  weak convergence holds.
\end{enumerate}

  We emphasize here the point that we would like
the reader to understand by a \emph{non-commutative
reading of the complex plane} \cite{Madore}: a non-commutative algebra of
operators in $\H$ takes the place of the commutative algebra of complex
valued functions on the plane. Note that the map (\ref{stquant}) is linear
and, in view of (\ref{stresunity}), the function $f(z,\bar{z}) =1$ is mapped to
the identity. These features are what we ``minimally'' expect from any
quantization scheme.

For the simplest functions $f(z,\bar{z}) = z$ and $f(z,\bar{z}) = \bar{z}$ we obtain
\begin{align}
\label{stoper1}
   A_z &= a\, ,  \quad a \, e_n = \sqrt{n} \, e_{n-1}\, , \quad a\,e_0 = 0 \, ,\quad
   \mbox{(lowering operator)}\\
   A_{\bar z} & = a^{\dag} \, , \quad a^{\dag} \, e_n = \sqrt{n +1} \, e_{n+1}
   \quad \mbox{(raising operator)}\, .
\end{align}
These two basic operators obey the so-called canonical commutation rule :
$[a,a^{\dag}] = \I$. The \emph{number operator} $N$ is defined by
$N = a^{\dag} a$ and is such that its spectrum is exactly $\N$ with
eigenvectors $e_n$ : $N e_n = n e_n$. The linear span of the triple
$\{a, a^{\dag}, \I\} $ equipped with the operator commutator $[\cdot\, , \cdot]$
is the \emph{Weyl-Heisenberg} Lie algebra, a key mathematical object for standard
quantum mechanics. The fact that the complex plane has become non-commutative is
apparent from the quantization of the real and imaginary parts of $z =
\dfrac{1}{\sqrt{2}}(q + ip)$ :
\begin{equation}
\label{stQP}
A_q \deq Q = \frac{1}{\sqrt{2}}(a  + a^{\dag})\, , \quad A_p \deq P =
\frac{1}{\sqrt{2}i}(a  - a^{\dag})\,, \quad [Q,P] = i\I\, .
\end{equation}

\section{Generic Delone sequences  and the ``fuzzy plane''}
\label{gendolfuzz}
We now consider a generic Delone sequence as in  (\ref{seq}). As a first remark, we easily assert the following result about the convergence of the associated exponential.

 \begin{Prop}
 Let $\mathcal{X} = \{ x_n\}_{n\in \N}$ be a non-negative sequence such that $x_n > 0$ for all $n \geq 1$ and $\lim_{n \to \infty} x_n = \infty$. Then its associated exponential
 \begin{equation}
\label{seqexp}
\mathcal{N}(t) = \sum_{n = 0}^{+\infty} \frac{t^n}{x_n!}\, ,
\end{equation}
has an infinite convergence radius.
\end{Prop}
\DEM
It is  enough to apply the d'Alembert criterion for a power series of the form $\sum_{n = 0}^{\infty} a_n z^n$ :
$$
\lim_{n \to \infty} \left\vert\frac{a_{n+1}}{a_n}\right\vert =\lim_{n \to \infty}
\frac{1}{x_{n+1}} = 0\, .
$$
\QED

We then mimic the construction (\ref{stcs}) by considering the following family
of vectors in
$\H$ :
\begin{equation}
\label{xncs}
v_z = \sum_{n=0}^{\infty} \frac{1}{\sqrt{\mathcal{N}(\vert z \vert^2)}}\frac{z^n}{\sqrt{x_n!}}\,  e_n\, .
\end{equation}
These vectors still enjoy  some properties similar to the standard ones.
\begin{itemize}
  \item[(i)] $\Vert v_z\Vert = 1$ (\textbf{normalization}).
  \item[(ii)] The map $\C \ni z \mapsto v_z$ is weakly
  continuous (\textbf{continuity}).
  \item[(iii)] The map $\N \in n \mapsto \vert (e_n\, ,v_z)\vert^2 =
  \dfrac{\vert z \vert^{2n}}{\mathcal{N}(\vert z \vert^2)x_n!}$ is a
  Poisson-like distribution in $\vert z\vert^2$,  with average number of successes equal
 to $\vert z \vert^2$
  (\textbf{discrete probabilistic content} \cite{alhelgaz,barcrow}).
  \end{itemize}
  However, the map $\C \ni z \mapsto \vert (e_n\, ,v_z)\vert^2 =
\dfrac{\vert z \vert^{2n}}{\mathcal{N}(\vert z \vert^2)x_n!}$ is not a (Gamma-like)
probability distribution (with respect to the square of the radial variable in the
complex plane) with $x_{n+1}$ as a shape parameter, and this is a serious setback
for the Berezin-Toeplitz quantization program. Indeed  there is no reason why
carrying out again  the calculation (\ref{stquant}) should yield the resolution of
the unity : with $\mathrm{P}_z = (v_z\, , \cdot)v_z$,
\begin{align}
\label{genquant}
\nonumber \int_{\C}\dfrac{d^2 z}{\pi}\,  \mathrm{P}_z &= \sum_{n, n'= 0}^{\infty}
e_n\,  (e_{n'}\, ,\cdot) \, \frac{1}{\sqrt{x_n!x_n'!}}\, \int_{\C}\dfrac{d^2 z}{\pi}\,
\frac{1}{\mathcal{N}(\vert z \vert^2)} z^n{\bar z}^{n'}\\ & =
\sum_{n= 0}^{\infty}\frac{1}{x_n!} \mathcal{I}(n)  e_n\,  (e_{n}\, ,\cdot) \deq
\mathbb{F}\,
\end{align}
where $\mathbb{F}$ is a diagonal operator determined by the sequence of integrals
\begin{equation}
\label{genIn}
\mathcal{I}(n) = \int_{0}^{+\infty} t^n \, \frac{dt}{\mathcal{N}(t)}\, .
\end{equation}
These integrals form a sequence of Stieltjes moments for the measure
$\dfrac{dt}{\mathcal{N}(t)}$.

We are now faced with the following alternatives.

\paragraph{The moment problem has a solution.}
More precisely,  we first suppose that the Stieltjes moment problem has a
solution for the sequence of factorials $\left(x_n!\right)_{n\in \N}$, i.e.
there exists a probability distribution $t \mapsto w(t)$  on $[0, +\infty)$
\cite{alhelgaz} with infinite support such that
\begin{equation}
\label{genmom}
x_n! = \int_{0}^{+\infty} t^n \, w(t)\, dt\, .
\end{equation}
We know that a necessary and sufficient condition for this is that   the two  matrices
\begin{equation}
\label{matmom}
\begin{pmatrix}
  1    & x_1! & x_2! & \dotso  &x_n!  \\
 x_1! & x_2! & x_3! & \dotso  &x_{n+1}!  \\
 x_2! & x_3! & x_4! & \dotso  &x_{n+2}!  \\
  \vdots & \vdots & \vdots & \ddots &  \vdots\\
  x_n! & x_{n+1}! & x_{n+2}!& \dotso & x_{2n}!
\end{pmatrix}\, , \quad
\begin{pmatrix}
 x_1! & x_2! &  x_3! & \dotso  &x_{n+1}!  \\
 x_2! & x_3! & x_4! & \dotso  &x_{n+2}!  \\
  x_3! & x_4! & x_5! & \dotso  &x_{n+3}!  \\
  \vdots & \vdots & \vdots & \ddots &  \vdots\\
 x_{n+1}! & x_{n+2}!& x_{n+3}!&\dotso & x_{2n+1}!
\end{pmatrix}
\end{equation}
have strictly positive determinants for all $n$.

Then, a  natural approach is just to modify the measure in (\ref{genmom})
by including the weight $w(\vert z \vert ^2)\,\mathcal{N}(\vert z \vert ^2)$.
We then obtain the resolution of the identity :

\begin{align}
\label{genquant1}
\nonumber \int_{\C}\dfrac{d^2 z}{\pi}\, w(\vert z \vert ^2)\,\mathcal{N}(\vert z
\vert ^2)\,  \mathrm{P}_z &= \sum_{n, n'= 0}^{\infty}  e_n\,  (e_{n'}\, ,\cdot) \,
\frac{1}{\sqrt{x_n!x_n'!}}\, \int_{\C}\dfrac{d^2 z}{\pi}\,
w(\vert z \vert ^2)\,z^n{\bar z}^{n'}\\ & =  \sum_{n= 0}^{\infty}\frac{1}{x_n!} \,
\int_{0}^{+\infty} t^n \, w(t)\, dt\,   e_n\,  (e_{n}\, ,\cdot) =\mathbb{I}\, .
\end{align}

Of course, in many (if not most) cases the explicit form of the measure
is not known. So, at this stage, the construction of CS may remain at a formal level
and we will rather adopt the approach described below.

\paragraph{The moment problem has no solution  or has a solution with unsolved measure.}

Here we will suppose that the sequence of ratios
\begin{equation}
\label{genrat1}
\mu_n \DEF \frac{\mathcal{I}(n)}{x_n!}
\end{equation}
has a finite limit as $n \to \infty$,
\begin{equation}
\label{genrat}
\mu_n  \to \mu_{\infty} < \infty\, \Leftrightarrow \, \lim_{n \to
\infty}\frac{\mu_n}{\mu_{n-1}} = 1\, .
\end{equation}

It is then natural to ``renormalize" the non-negative Delone sequence
(\ref{seq}) as follows.
\begin{equation}
\label{seqren}
\tilde{x}_n \DEF \frac{\mu_n}{\mu_{n-1}}\, x_n\,, \quad n \geq 1\, , \quad \tilde{x}_0  = 0\, ,
\end{equation}
and this entails
\begin{equation}
\label{seqren1}
\tilde{x}_n! = \frac{\mu_n}{\mu_0}\, (x_n!)\,, \quad n \in \N\, .
\end{equation}
The assumption (\ref{genrat})  ensures that this
renormalized sequence gets closer and closer to the original sequence, and
the condition of the terms being strictly
increasing is fulfilled beyond a certain point $n_0$.  If some crossing
happens for the first few elements of the sequence, we will reorder the sequence
(\ref{seqren}) so that it becomes  strictly increasing;
however, in order to avoid
an inflation of notations, we shall maintain the same notation (possibly
at the expense of some notational abuse).

We then start again the construction in (\ref{stcs}) by considering this time  the
renormalized family of vectors in
$\H$ :
\begin{equation}
\label{renxncs}
\tilde{v}_z = \sum_{n=0}^{\infty} \frac{1}{\sqrt{\widetilde{\mathcal{N}}(\vert z
\vert^2)}}\frac{z^n}{\sqrt{\tilde{x}_n!}}\,  e_n\, ,
\end{equation}
where $\widetilde{\mathcal{N}}(t)$ is the exponential associated to the new sequence,
\begin{equation}
\label{renexpo}
\widetilde{\mathcal{N}}(t) = \sum_{n = 0}^{+\infty} \frac{t^n}{\tilde{x}_n!}\, .
\end{equation}
From (\ref{genrat}) it is clear that this series also has an infinite
radius of convergence.
Now, not only do the vectors (\ref{renxncs}) enjoy  properties similar to the
vectors (\ref{xncs}),
\begin{itemize}
  \item[($\tilde{i}$)] $\Vert \tilde{v}_z\Vert = 1$ (\textbf{normalization}).
  \item[($\widetilde{ii}$)] The map $\C \ni z \mapsto \tilde{v}_z$ is continuous
  (\textbf{continuity}).
  \item[($\widetilde{iii}$)] The map $\N \in n \mapsto \vert (e_n\, ,\tilde{v}_z)
  \vert^2 =  \dfrac{\vert z \vert^{2n}}{\widetilde{\mathcal{N}}(\vert z \vert^2)
  \tilde{x}_n!}$ is a Poisson-like distribution with average number of occurrences
  equal to $\vert z \vert^2$  (\textbf{discrete probabilistic content}),
  \end{itemize}
but they also enjoy what we need for carrying out a  Berezin-Toeplitz like
quantization, since :
\begin{itemize}
\item[($\widetilde{iv}$)] The map $\C \ni z \mapsto \vert (e_n\, ,\tilde{v}_z)
\vert^2 = \dfrac{\vert z \vert^{2n}}{\widetilde{\mathcal{N}}(\vert z \vert^2)
\tilde{x}_n!}$ \underline{is} a (Gamma-like) probability distribution (with
respect to the square of the radial variable) with $\tilde{x}_{n+1}$ as a
shape parameter \underline{and} with respect to the modified measure on the
complex plane
\begin{equation}
\label{modmeas}
\nu(dz) \DEF \frac{\widetilde{\mathcal{N}}(\vert z \vert^2)}{\mathcal{N}(\vert z
\vert^2)}\, \frac{d^2 z}{\mu_0\, \pi}\, .
\end{equation}
\item[($\widetilde{v}$)] The family of vectors (\ref{renxncs})  solves  the unity :
with $\widetilde{\mathrm{P}}_z = (\tilde{v}_z\, , \cdot)\tilde{v}_z$,
\begin{align}
\label{rengenquant}
\nonumber \int_{\C}\nu(dz) \,  \widetilde{\mathrm{P}}_z &= \sum_{n, n'= 0}^{\infty}  e_n\,  (e_{n'}\, ,\cdot) \, \frac{1}{\sqrt{\tilde{x}_n!\tilde{x}_n'!}}\, \int_{\C}\dfrac{d^2 z}{\mu_0\, \pi}\,  \frac{1}{\mathcal{N}(\vert z \vert^2)} z^n{\bar z}^{n'}\\ & =  \sum_{n= 0}^{\infty}\frac{1}{\tilde{x}_n!} \mathcal{I}(n)  e_n\,  (e_{n}\, ,\cdot) = \mathbb{I}\, ,
\end{align}
the last equality resulting trivially from (\ref{genrat}) and (\ref{seqren1}).

\end{itemize}

\paragraph{A statistical interpretation of the measure $\nu(dz)$.}
As we already noted, the map $ \N \ni n \mapsto \frac{1}{\mathcal{N}(t)}\, \frac{t^n}{x_n!}$ is a Poisson-like distribution determined by the original sequence $\mathcal{X} = (x_n)$. On the other hand, the map
\begin{equation}
\label{maprdvar}
\N \ni n \mapsto \sigma(n) \deq \frac{1}{\mu_n}
\end{equation}
defines the random variable $\sigma(n)$. 
Hence, the ratio $\widetilde{\mathcal{N}}(t)/\mu_0 \mathcal{N}(t)$ is nothing else that the expectation value of $\sigma$ with respect to the Poisson-like distribution defined by $\mathcal{X}$:
\begin{equation}
\label{expsigma}
\frac{\widetilde{\mathcal{N}}(t)}{\mu_0\,\mathcal{N}(t)} = \sum_{n=0}^{\infty} \frac{1}{\mu_n}\, \frac{1}{\mathcal{N}(t)}\, \frac{t^n}{x_n!} \equiv (\mathrm{E}_{\mathcal{X}}\, \sigma)(t)\, .
\end{equation}
In consequence, the measure $\nu(dz)$ on the complex plane reads as 
\begin{equation}
\label{nuexpect}
\nu(dz) = (\mathrm{E}_{\mathcal{X}}\, \sigma)(\vert z \vert^2)\, \frac{d^2z}{\pi}\, .
\end{equation}

\paragraph{A comment regarding the ``perturbed'' Lebesgue measure $\nu(dz)$
in (\ref{modmeas}) in the case where the moment problem has an explicit  solution
with  resolving measure $w(t)\, dt $}:\\

Even in this case, where an explicit solution to the moment problem  (\ref{genmom})
exists, it is interesting to follow the second route: keeping the perturbed
Lebesgue measure (\ref{modmeas}) in the plane,  carrying out  the alternate quantization
using it  and
comparing it with the quantization of the phase space $\C$ equipped with the
{\em non-symplectic measure} $w(\vert z \vert^2)\, \mathcal{N}(\vert z \vert^2)\, d^2z/\pi$.  We will examine in Section \ref{genhyp} this question
in the simplest example of the constant shift $\alpha(n) = \epsilon$, $n \geq 1$.
We will also discuss in the conclusion the relevance of such different  approaches to the ``quantum processing'' of the  classical phase space, provided with the Lebesgue measure
(for which classical states are just point or Dirac measures), or alternatively,
 provided with a  different measure, giving a ``mixed'' nature to classical states.

 Note also, that in the case which concerns us most in this
 paper, of $\alpha$-perturbations of the natural numbers, since $\mu_n / \mu_{n-1}
 \rightarrow 1$, as $n \rightarrow \infty$ (see Eq. (\ref{genrat})),  the discrete distribution,
 $n \mapsto ( \widetilde{\mathcal N} (\vert z \vert^2  ))^{-1}  \vert z \vert^{2n}/\widetilde{x}_n !$
 appears as a
 ``perturbation'' of the Poisson distribution. Similarly the continuous distribution
 $\vert z \vert^2 \mapsto   (\widetilde{\mathcal N} \vert z \vert^2 ) )^{-1}  \vert z \vert^{2n}/\widetilde{x}_n !$ can
 be looked upon as a ``perturbed'' Gamma-distribution. In this sense we also have here
 a ``mild perturbation'' on the original Bayesian duality, involving the Poisson and
 Gamma-distributions characterizing  the case of the standard coherent states.


\section{Quantization of the classical phase space through renormalized Delone sequences}

We know that Property ($\widetilde{v}$) is crucial since it allows to define

\begin{enumerate}
  \item[($\widetilde{vi}$)] a normalized positive operator-valued measure
  (POVM) on the complex plane equipped 
  with its Lebesgue measure
  $\dfrac{d^2 z}{\pi}$ and its $\sigma-$algebra $\mathcal{F}$ of Borel sets :
\begin{equation}
\label{stpovnm}
\mathcal{F} \ni \Delta \, \mapsto \int_{\Delta}\frac{\widetilde{\mathcal{N}}(\vert z \vert^2)}
{\mathcal{N}(\vert z \vert^2)}\,\dfrac{d^2 z}{\mu_0\, \pi}\,
\mathrm{\widetilde{P}}_z \, \in \mathcal{L}(\mathcal{H})^+\, ,
\end{equation}
where $\mathcal{L}(\mathcal{H})^+$ is the cone of positive bounded operators on $\H$;
  \item[($\widetilde{vii}$)] the Berezin-Klauder-Toeplitz quantization of the complex plane :
  \begin{equation}
\label{stquantnm}
f \mapsto \widetilde{A}_f = \int_{\C} \frac{\widetilde{\mathcal{N}}(\vert z \vert^2)}
{\mathcal{N}(\vert z \vert^2)}\,\dfrac{d^2 z}{\mu_0\,\pi}\, f(z,\bar{z}) \, \mathrm{\widetilde{P}}_z =
 \sum_{n, n'= 0}^{\infty}  (\widetilde{A}_f)_{nn'}\, e_n\,  (e_{n'}\, ,\cdot) \, ,
\end{equation}
where
\begin{align}
\label{matelnm}
\nonumber (\widetilde{A}_f)_{nn'}&= \frac{1}{\sqrt{\tilde{x}_n!\tilde{x}_{n'}!}}\, \int_{\C}
\frac{1}{\mathcal{N}(\vert z \vert^2)}\,
\dfrac{d^2 z}{\mu_0\,\pi}\, f(z,\bar{z}) z^n{\bar z}^{n'}\\ &= \frac{1}{\sqrt{\tilde{x}_n!\tilde{x}_{n'}!}}\,
\int_0^{\infty}\frac{du}{\mu_0\mathcal{N}(u)}\,u^{\frac{n+n'}{2}}\,
\left\lbrack\dfrac{1}{2\pi}\int_0^{2\pi}d\theta\, e^{-i(n'-n)\theta}\,
f(\sqrt{u}e^{i\theta}, \sqrt{u}e^{-i\theta}) \right\rbrack\, ,
\end{align}
 provided that  weak convergence holds.
\end{enumerate}

For the simplest functions $f(z,\bar{z}) = z$ and $f(z,\bar{z}) = \bar{z}$ we obtain
lowering and raising operators respectively,
\begin{align}
\label{stoper}
   \widetilde{A}_z &= \tilde{a}\, ,  \quad \tilde{a} \, e_n = \sqrt{\tilde{x}_n} e_{n-1}\, ,
   \quad a\,e_0 = 0 \, ,\quad \mbox{(lowering operator)}\\
    \widetilde{A}_{\bar z} & = \tilde{a}^{\dag} \, , \quad \tilde{a}^{\dag} \, e_n =
     \sqrt{\tilde{x}_{n+1}} e_{n+1} \quad \mbox{(raising operator)}\, .
\end{align}
These operators obey the non-canonical commutation rule : $[\tilde{a},\tilde{a}^{\dag}]
= \tilde{x}_{N+1} - \tilde{x}_{N}$, where $\tilde{x}_{N}$ is defined by
$\tilde{x}_{N} = \tilde{a}^{\dag} \tilde{a}$ and is such that its spectrum is
exactly $\{\tilde{x}_n\, , \, n \in \N\}$ with eigenvectors $\tilde{x}_{N}\,e_n=
\tilde{x}_n e_n$. The linear span of the triple $\{\tilde{a}, \tilde{a}^{\dag},
\widetilde{X}_N\} $  is obviously not closed, in general,  under commutation and the set
of resulting commutators  generically gives  rise to an infinite
dimensional Lie algebra.

The next simplest function on the complex plane is the squared modulus  
of $z$, $\vert z \vert^2 = \frac{1}{2}(p^2 + q^2)$, i.e. 
one-half times the squared radius of the circle, in other 
words, the classical Hamiltonian of the harmonic oscillator 
when we consider the complex plane as the phase space of a
particle moving on the line:
\begin{equation}
\label{qcsho}
\widetilde{A}_{z\bar z} = \tilde{a}\, \tilde{a}^{\dag} =\tilde x_{N+1}\, .
\end{equation}
Therefore, the spectrum of the quantized version of $\vert z \vert^2 $  is the 
renormalized sequence (\ref{seqren}) up to a shift by 1. We will come back 
to this important fact in the conclusion.

\section{Localization properties and associated orthogonal polynomials}
Consider the operators:
\begin{equation}
 \widetilde{Q} = \frac 1{\sqrt{2}}[\tilde{a} + \tilde{a}^\dagger]\; , \qquad
 \widetilde{P} = \frac 1{i\sqrt{2}}[\tilde{a} - \tilde{a}^\dagger]\; .
\end{equation}
The operator $\widetilde{Q}$ acts on the basis vectors $e_n$ in the manner,
\begin{equation}
  \widetilde{Q} e_n = \sqrt{\frac{\tilde{x}_n}2}\; e_{n-1} +
  \sqrt{\frac{\tilde{x}_{n+1}}2}\; e_{n+1}\; .
\label{eq:pos-op-act}
\end{equation}
Since for a Delone sequence the sum $\displaystyle{\sum_{n=0}^\infty
\frac 1{\sqrt{\tilde{x}_n}}}$ diverges, the operator $\widetilde{Q}$ is essentially
self-adjoint \cite{borzov} and hence has a 
unique self-adjoint extension, which we again
denote by $\widetilde{Q}$. Let $E_\lambda , \;
 \lambda \in \mathbb R$, be the spectral family of $\widetilde{Q}$, so that,
$$ \widetilde{Q} = \int_{-\infty}^\infty \lambda \; dE_\lambda \; .$$
Thus there is a measure $ \varpi(d\lambda)$ on $\mathbb R$ such that on the
Hilbert space $L^2 (\mathbb R , \varpi)$,
the action of
$\widetilde{Q}$ is just a multiplication by $\lambda$ and the basis vectors
$e_n$ can  be represented by functions
$p_n (\lambda )$. Consequently, on this space, the relation (\ref{eq:pos-op-act})
assumes the form
\begin{equation}
  \lambda p_n (\lambda) = c_n  p_{n-1}(\lambda ) + c_{n+1} p_{n+1} (\lambda)\; ,
   \qquad c_n =  \sqrt{\frac{\tilde{x}_n}2}\;,
\label{eq:pos-op-act2}
\end{equation}
which is a two-term recursion relation, familiar from the theory of orthogonal
polynomials. It follows that
$\varpi (d\lambda ) = d(e_0, E_\lambda e_0) $, and the $p_n$
may be realized as the polynomials obtained
by orthonormalizing the sequence of monomials $1, \lambda, \lambda^2 , \lambda^3 ,
\ldots\; , $ with respect to this measure
(using a Gram-Schmidt procedure).  Furthermore, for any $\varpi$-measurable set
$\Delta \subset \mathbb R$,
\begin{equation}
(e_n, E(\Delta)e_m) = \int_{\Delta}  p_n (\lambda )p_m (\lambda )\;\varpi (d\lambda )\; ,
\label{eq:poly-basis}
\end{equation}
and
\begin{equation}
   (p_n, p_m)_{L^2 (\mathbb R , \varpi )} = \int_{\mathbb R}
       p_n (\lambda )p_m (\lambda )\; \varpi (d\lambda ) = \delta_{m n}\; .
\label{eq:poly-orthog}
\end{equation}

The polynomials $p_n$ are not {\em monic polynomials\/}, i.e., that the coefficient of $\lambda^n$ in
$p_n$ is not one. However, the renormalized polynomials
\begin{equation}
  q_n (\lambda ) = {c_n !}\; p_n (\lambda ) \, ,
\end{equation}
are seen to satisfy the recursion relation
\begin{equation}
  q_{n + 1} (\lambda ) = \lambda\; q_n (\lambda )  - c_n^2\; q_{n-1} (\lambda )\; ,
\label{eq:mon-rec-reln}
\end{equation}
from which it is clear that these polynomials are indeed monic.
Also, writing $\zeta_z$ for the vector $\widetilde{\mathcal N}(\vert z \vert^2 )^{\frac 12}\; \tilde{v}_z$, when
expressed as a function in $L^2 (\mathbb R , \varpi )$, we see that the function
\begin{equation}
 G(z, \lambda ) :=  \zeta_z (\lambda ) =  \sum_{n=0}^\infty \frac {z^n}{\sqrt{\tilde{x}_n !}}\; p_n (\lambda )
  = \sum_{n=0}^\infty \frac {2^{\frac n2}z^n}{\tilde{x}_n !}\; q_n (\lambda ) \; ,
\label{eq:orth-polynCS}
\end{equation}
is the generating function for the polynomials $q_n$ (or $p_n$), in the sense that
\begin{equation}
   q_n (\lambda ) = \frac 1{2^{\frac n2}}\;\frac {\tilde{x}_n !}{n!} \;\left.\frac {\partial^n G (z, \lambda )}
       {\partial z^n} \right|_{z=0}\; .
\end{equation}

   There is a simple way to compute the monic polynomials. To see this, note first that in virtue of
(\ref{eq:pos-op-act}) and (\ref{eq:pos-op-act2}), the operator $\widetilde{Q}$ is represented in the $e_n$
basis as the infinite tri-diagonal matrix,
\begin{equation}
\widetilde{Q} = \begin{pmatrix} 0 & c_1 & 0 & 0 & 0 &\ldots \\ c_1 & 0 & c_2 & 0 & 0 & \ldots\\
0 & c_2 & 0 & c_3 & 0 & \ldots\\
0 & 0 & c_3 & 0 & c_4 & \ldots\\0 & 0 & 0 & c_4 & 0 & \ldots\\
\vdots & \vdots &\vdots &\vdots &\vdots &\ddots \end{pmatrix}\; .
\end{equation}
Let $\widetilde{Q}_n$ be the truncated matrix consisting of the first $n$ rows and columns of $\widetilde{Q}$  and $\mathbb I_n$
the $n\times n$ identity matrix. Then,
\begin{equation}
  \lambda \mathbb I_n - \widetilde{Q}_n =
\begin{pmatrix} \lambda &  - c_1 & 0 & 0 & 0 &\ldots & 0 & 0 & 0\\
- c_1 & \lambda & - c_2 & 0 & 0 & \ldots & 0 & 0 & 0\\
0 & - c_2 & \lambda & - c_3 & 0 & \ldots & 0 & 0 & 0\\
0 & 0 & - c_3 & \lambda & - c_4 & \ldots & 0 & 0 & 0 \\
0 & 0 & 0 & - c_4 & \lambda & \ldots & 0 &0 & 0\\
\vdots & \vdots &\vdots &\vdots &\vdots &\ddots & \vdots & \vdots & \vdots\\
0 & 0 & 0 & 0 & 0 & \ldots & \lambda & -c_{n-2} & 0\\
0 & 0 & 0 & 0 & 0 & \ldots & -c_{n-2} & \lambda & -c_{n-1}\\
0 & 0 & 0 & 0 & 0 & \ldots & 0 & - c_{n-1} & \lambda \end{pmatrix}\; .
\end{equation}
It now follows that $q_n$ is just the characteristic polynomial of $\widetilde{Q}_n$ :
\begin{equation}
  q_n (\lambda ) = \text{det} [ \lambda \mathbb I_n - \widetilde{Q}_n ]\; .
\end{equation}
Indeed, expanding the determinant with respect to the last row, starting at the lower right
corner, we easily get
\begin{equation}
\text{det} [ \lambda \mathbb I_n - \widetilde{Q}_n ] = \lambda \;\text{det} [ \lambda \mathbb I_{n-1} - \widetilde{Q}_{n-1} ] -
    c_{n-1}^2 \;\text{det} [ \lambda \mathbb I_{n-2} - \widetilde{Q}_{n-2} ]\; ,
\end{equation}
which is precisely the recursion relation (\ref{eq:mon-rec-reln}). Consequently the roots of the polynomial
$q_n$ (or $p_n$) are the eigenvalues of $\widetilde{Q}_n$.

\section{Localization properties of the quantum phase space viewed through lower symbols}

Given a function $f$ on the  complex plane, the resulting operator $\widetilde{A}_f$,
if it exists, at least  in a weak sense, acts on the Hilbert space ${\mathcal H}$
with orthonormal basis $e_n$ : the  integral
\begin{equation}
\label{wksssymb}
(\psi, \widetilde{A}_f \psi)= \int_{\C}f(z,\bar{z})\vert (\psi, \tilde{v}_z)\vert^2  \,
\nu(dz)
\end{equation}
should be finite for all  $\psi$ in some dense subset  of
$\mathcal{H}$. It should be noted that  if $\psi$ is normalized then
(\ref{wksssymb}) represents the mean value of the function $f$ with respect
to the $\psi$-dependent probability distribution $z \mapsto \vert
(\psi, \tilde{v}_z)\vert^2$ on the complex plane.

In order to be more rigorous on this important point, let us adopt the following
acceptance criteria for a function  to belong to the class of quantizable classical
observables.

\begin{defi}
\label{defobscs}
A  function $\C \ni z \mapsto f(z,\bar{z}) \in \C$  is a \emph{CS quantizable
classical observable} via the  map $f \mapsto \widetilde{A}_f$ defined by
(\ref{stquantnm})
if the map $\C \ni z = \frac{1}{2} (q + ip) \equiv (q,p) \mapsto (\tilde{v}_z,
\widetilde{A}_f \tilde{v}_z)$  is a smooth (i.e. $ \in C^{\infty}$) function
with respect to the $(q,p)$ coordinates of the complex plane.
 \end{defi}

 The function $f$  is the \emph{upper} \cite{csfks} or \emph{contravariant}
 \cite{ber} symbol of the operator $\widetilde{A}_f$, and $ (\tilde{v}_z ,
 \widetilde{A}_f \tilde{v}_z)$
 is the \emph{lower} \cite{csfks} or \emph{covariant}
 \cite{ber} symbol of the operator $\widetilde{A}_f$.

In \cite{chagayou} such a definition is extended to a class of  distributions including tempered distributions.

Hence,  localization properties in the complex plane, from the point of view of the
sequence $\{ x_n \}_{n\in \N}$ and its renormalized
version $\{ \tilde{x}_n \}_{n\in \N}$,
should be examined through the shape (as functions of  $z$) of the respective
lower symbols
\begin{equation}
\label{lowsymbQP}
\check{\tilde{Q}}(z) \deq (\tilde{v}_z,   \widetilde{Q} \tilde{v}_z)\, ,
\quad \check{\tilde{P}}(z) \deq (\tilde{v}_z,   \widetilde{P} \tilde{v}_z)\, ,
\end{equation}
and the noncommutativity of the reading of the complex plane should be
encoded in the behavior of the lower symbol  $(\tilde{v}_z, [ \widetilde{Q},
\widetilde{P}] \tilde{v}_z)$ of the commutator $[\widetilde{Q}, \widetilde{P}]$.
The study, within the above framework, of the product of dispersions
\begin{equation}
\label{uncert}
\left(\Delta_{\tilde{v}_z}\widetilde{Q}\right)\, \left(\Delta_{\tilde{v}_z} \widetilde{P}\right) = \frac{1}{2}\, (v_z, (\tilde{x}_{N+1} - \tilde{x}_N)v_z)  \equiv  \frac{1}{2}\, (\tilde{v}_z, \tilde{x}'_{N}\, \tilde{v}_z)
\end{equation}
expressed in states $\tilde{v}_z$  should thus be relevant. Note that in the case of an $\alpha$-perturbation of  $\N$, the r.h.s of this equality becomes $ \frac{1}{2}\,(1 +  (\tilde{v}_z, \tilde{\alpha}'(N)\,\tilde{v}_z)$, where $ \tilde{\alpha}_n \deq \tilde{x}_n - n$. Introducing supremum $s_{\tilde{\alpha}}$ and infimum $i_{\tilde{\alpha}}$ for the set of ``renormalized'' fluctuations $\{\tilde{\alpha}_{n+1} - \tilde{\alpha}_n\, , \, n\in \N\}$,
we get  the following bounds on the product of dispersions :
\begin{equation}
\label{}
  \frac{1}{2}\,(1 +  i_{\tilde{\alpha}})\leq  \left(\Delta_{\tilde{v}_z}\widetilde{Q}\right)\, \left(\Delta_{\tilde{v}_z} \widetilde{P} \right)\leq   \frac{1}{2}\,(1 +  s_{\tilde{\alpha}})\, .
\end{equation}

\section{Delone sequences and generalized hypergeometric functions}
\subsection{A statistical perturbation of the Bernoulli process}
It is well known that the Poisson distribution $\N \ni k \mapsto e^{- t}t^k/k!$
with parameter $t$ is the  large $n$ limit of the Bernouilli process of a sequence
of $n$ trials with two possible outcomes, one (``win'', say) with probability $p = t/n$,
the other one (``loss'') with probability  $q = 1-p$. The probability  to get $k$ wins
after $n$ trials is given by  the binomial distribution :
\begin{equation}
\label{bern1}
 p_k^{(n)} = \binom{n}{k}\, p^k \, (1-p)^{n-k}\, ,
\end{equation}
and its Poisson limit is readily obtained. Now, imagine a sequence of $n$ trials
for which the probability  to obtain  $k$ wins is given by the following ``perturbation''
of the  binomial distribution :
\begin{equation}
\label{berndef}
 \mathfrak{p}_k^{(n)} = \frac{x_n!}{x_{n-k}! \, x_k!}\, p^k \, q_{n;k}(p)\, ,
 \quad \sum_{k=0}^n \mathfrak{p}_k^{(n)}(p)=1\, ,
\end{equation}
where $\{x_n, \, n \in \N\}$ is an $\alpha$-Delone perturbation of the natural
numbers. Here, the sequence of  unknown functions $q_{n;k}(p)$ should be
such that, in the limit $n \to \infty$ with  $p = t/x_n$, we get our deformation
of the Poisson law:
\begin{equation}
\label{poisslim}
 \mathfrak{p}_k^{(n)} = \frac{x_n!}{x_{n-k}! \, x_k!}\, p^k \, q_{n;k}(p)
 \underset{n \to \infty}{\to} \frac{1}{\mathcal{N}(t)}\, \frac{t^k}{x_k!}\, .
\end{equation}
In order to have this limit, we first observe with $p= t/x_n$ that for an
$\alpha$-Delone perturbation
\begin{equation}
\label{poisslim1}
\frac{x_n!}{x_{n-k}! \, x_k!}\, p^k \, q_{n;k}(p) =\frac{x_n}{x_n}\,
\frac{x_{n-1}}{x_n}\dotsb \frac{x_{n-k + 1}}{x_n}\, \frac{t^k}{x_k!}\,
q_{n;k}\left(\frac{t}{x_n}\right) \underset{n \to \infty}{\to} \frac{t^k}{x_k!}
\underset{n \to \infty}{\lim} q_{n;k}\left(\frac{t}{x_n}\right)\, ,
\end{equation}
and so we should have
\begin{equation}
\label{poisslim2}
\underset{n \to \infty}{\lim} q_{n;k}\left(\frac{t}{x_n}\right) =
\frac{1}{\mathcal{N}(t)}\, .
\end{equation}

\subsection{The generalized hypergeometric function as an approximation of the
exponential}
\label{genhyp}
In view of Eq. \ref{poisslim2}, we would like to find an example of a perturbation
which yields for the
series $\mathcal{N}(t)$ something close to the ordinary exponential. A good
candidate for this is precisely the confluent hypergeometric or Kummer function :
\begin{align}
\label{conf}
\nonumber \mathcal{N}(t) &:= {}_{1}F_1(a; a + \epsilon; t) =\sum_{n=0}^{\infty}
\frac{a(a+1)\dotsb(a + n-1)}{(a + \epsilon)(a + \epsilon+1)\dotsb(a  +
\epsilon + n-1)} \, \frac{t^n}{n!} \\ &\equiv \sum_{n=0}^{\infty}
\frac{t^n}{x_n!}\, ,
\end{align}
where
\begin{equation}
\label{conf1}
x_n = n + \alpha(n) \equiv n  + \frac{\epsilon}{1 + \dfrac{a-1}{n}}\, ,
\end{equation}
and where $\epsilon$ is a small real parameter.
We thus get an $\alpha$-Delone perturbation of natural numbers which
is of the type of the  projective transformation (\ref{homseq}).
Moreover, we readily go smoothly back to the exponential as $\epsilon \to 0$.

Now, from the integral representation of the Kummer function,
\begin{equation}
\label{intkum}
{}_{1}F_1(a; a + \epsilon; t) = \frac{\Gamma(a + \epsilon)}{\Gamma(a)
\Gamma(\epsilon)}\, \int_0^1 e^{t\,  u} \, u^{a-1}\, (1-u)^{\epsilon -1}\, du\, ,
\end{equation}
and from the usual polynomial approximation of the exponential, we derive the
following binomial like approximation,
\begin{align}
\label{intkum1}
\nonumber {}_{1}F_1(a; a + \epsilon; t) &= \underset{n \to
\infty}{\lim}\frac{\Gamma(a + \epsilon)}{\Gamma(a) \Gamma(\epsilon)}\,
\int_0^1 \left(1 - \frac{t}{x_n}\, u \right)^{-x_{n-k}} \, u^{a-1}\,
(1-u)^{\epsilon -1}\, du\\
& =  \underset{n \to \infty}{\lim} {}_{2}F_1(x_{n-k},a;a + \epsilon; t)\, .
\end{align}
Hence, we get the explicit form for the functions $q_{n;k}(p)$:
\begin{equation}
\label{bern3}
q_{n;k}(p) = \frac{1}{\mathfrak{w}_n}\, \frac{1}{{}_{2}F_1(x_{n-k},a;a +
\epsilon; t)}\,
\end{equation}
where the normalization factor $ \dfrac{1}{\mathfrak{w}_n}$ is
defined by the condition
\begin{equation}
\label{normbern}
\sum_{k=0}^n \mathfrak{p}_k^{(n)}(p)=1\, .
\end{equation}

The simplest particular case of  the $x_n$ given by (\ref{conf1}), which
gives a perturbation of the Bernoulli process, is for $a = 1$, i.e. $x_n = n + \epsilon$, $n \geq 1$, $x_0 = 0$. In this
case, we are able to solve the moment problem (\ref{genmom}) with
\begin{equation}
\label{weight}
w(t) = \frac{t^\epsilon e^{-t}}{\Gamma(\epsilon + 1)} \, .
\end{equation}
Now, following the second way by studying the sequence of moment integrals given by
\begin{equation}
\label{mu}
\mu_n = \frac{1}{x_n!} \int _0^{\infty} \frac{t^n dt}{\NN(t)} ,
\end{equation}
with
\begin{equation}
\label{xisnfat}
x_n! = \frac{\Gamma(n + \epsilon + 1)}{\Gamma(\epsilon + 1)} \,  ,
\end{equation}
$\NN(t)$ is  straightforwardly calculated to be
\begin{equation}
\label{ene}
\NN(t) = \, _1F_1(1; 1 + \epsilon; t) = e^{t} t^{-\epsilon} [\Gamma(1 +
\epsilon) - \epsilon \Gamma(\epsilon, t)] \, ,
\end{equation}
where $\Gamma(\epsilon, t)$ is the incomplete Gamma function $\Gamma(\epsilon , t)
= \int_t^{\infty} x^{\epsilon-1} \exp(-x) dx$.
Then
\begin{equation}
\label{mun1}
\mu_n = \frac{\Gamma(\epsilon + 1)}{\Gamma(n + \epsilon + 1)} \int _0^{\infty}
\frac{t^{n+ \epsilon} e^{-t} dt}{\Gamma(\epsilon + 1) -
\epsilon \Gamma(\epsilon, t)} \,  .
\end{equation}

\begin{figure}[h]
\centering
\includegraphics[width=4.5in]{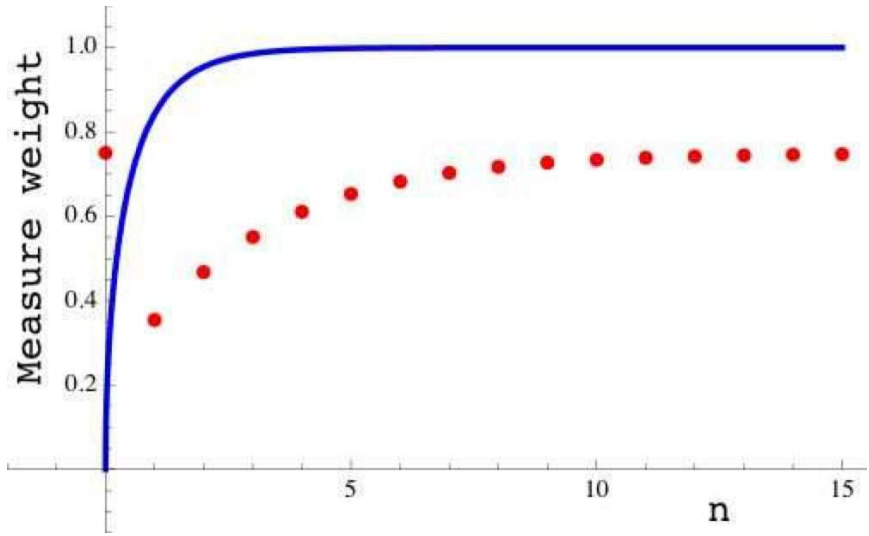}
\caption{The two measures for the sequence $x_n = n + \epsilon$, $n \geq 1$, and $x_0 = 0$ are compared through the  behavior of the two functions $ t \mapsto \left(1-\epsilon\frac{\Gamma(\epsilon,t)}{\Gamma(1 + \epsilon)}\right)$ (continuous line) and $ t \mapsto \frac{\widetilde{\mathcal{N}}(t)}{\mu_0\, \mathcal{N}(t)}$ (dot line) for $t \geq 0$. The value of $\epsilon$ has been chosen $\epsilon= 0.5$. }
\label{compmeas}
\end{figure}

As pointed out in the comment ending Section \ref{gendolfuzz}, we have here an elementary example allowing us to compare two types of non-commutaive reading of the complex plane. The first one stems from the solution of the Stieltjes moment problem and leads to the resolution of the identity with the measure 
\begin{equation}
\label{momeps1}
w(\vert z \vert^2) \, \mathcal{N}(\vert z \vert^2)\, \frac{d^2 z}{\pi}= \left(1-\epsilon\frac{\Gamma(\epsilon,\vert z \vert^2)}{\Gamma(1 + \epsilon)}\right)\, \frac{d^2 z}{\pi}\,.
\end{equation}
Note that in the limit $ z \to 0$ this measure reduces to $(1 - \epsilon)\, \frac{d^2 z}{\pi}$.
The second one is the measure associated with the renormalized sequence:
\begin{equation}
\label{momeps2}
\nu(dz) = \frac{\widetilde{\mathcal{N}}(\vert z \vert^2)}{\mathcal{N}(\vert z \vert^2)}\, \frac{d^2 z}{\mu_0\, \pi} = e^{-t} t^{\epsilon}\, \frac{\widetilde{\mathcal{N}}(\vert z \vert^2)}{ \Gamma(1 +
\epsilon) - \epsilon \Gamma(\epsilon, t)}\, \frac{d^2 z}{\mu_0\, \pi} \, ,
\end{equation} 
where the  $\widetilde{\mathcal{N}}(\vert z \vert^2)$ is numerically derived from the renormalized sequence	$(\tilde{x}_n)$. In Figure \ref{compmeas} we compare the two measures through the  behavior of the two functions $ t \mapsto \left(1-\epsilon\frac{\Gamma(\epsilon,t)}{\Gamma(1 + \epsilon)}\right)$ and $ t \mapsto \frac{\widetilde{\mathcal{N}}(t)}{\mu_0\, \mathcal{N}(t)}$ for $t \geq 0$. It is important to notice that the measure obtained by the solution of the moment problem
implies a radical change in the statistical content of the complex plane as a phase space, since at $t = 0$ the weight $w(t) \mathcal{N} (t)$ vanishes  for $\epsilon > 0$ and worse, it diverges for $\epsilon < 0$; in both cases  it tends to 1 as $t$ increases more or less rapidly depending on the value of $\epsilon$. On the other hand, the measure $\nu$ issued from the renormalized sequence has a more reasonable behavior, whatever the sign of $\epsilon$, since it starts at $1/\mu_0$ at $t = 0$, remains strictly positive,  and tends to $1/\mu_0$ at large $t$, more or less rapidly depending on $\vert \epsilon\vert$.

A second example which is interesting to study is the particular case
of (\ref{homseq}) for which $a = d = 0$ and $b/c = -\epsilon$, where
$\epsilon$ is a small real parameter. In that case,
\begin{equation}
\label{xisn2}
x_n = n(1 - \frac{\epsilon}{n^2})\, ,  \quad n \geq 1\, , \ x_0 = 0\, ,
\end{equation}

\begin{equation}
\label{xisn2fat}
x_n! = \frac{\Gamma(1-\sqrt{e}+n)\, \Gamma(1+\sqrt{e}+n)}{\Gamma(1-\sqrt{e}) \,
\Gamma(1+
\sqrt{e}) \,  n!} \,
\end{equation}
and
\begin{equation}
\label{2f2}
\NN(t) = \, \,  _2F_2 (1,1; 1-\sqrt{e},1+\sqrt{e};t) \, ,
\end{equation}
where $_2F_2(a_1,a_2;b_1,b_2;t)$ is a generalized hypergeometric  function.

As a matter of fact, the hypergeometric functions $ _1F_1(1; b; t)$ and
$_2F_2(1, 1; b_1,b_2; t)$ are particular cases of a more general class of
functions which also yield perturbations of the exponential function, that
is, $_qF_q(1_1, 1_2, ...1_q; b_1, b_2, ...b_q; t)$, that are connected
to Delone sequences of the form $x_n = n + \alpha^{(q)}(n)$, with
\begin{equation}
\label{alphagen}
\alpha^{(q)}(n) = \frac{\prod_{i = 1}^q (n - b_i) - n^q}{n^{q-1}}\,\, .
\end{equation}
For these sequences,
\begin{equation}
\label{xngen}
x_n! = \frac{\prod_{i=1}^q \frac{\Gamma(1 - b_i + n)}{\Gamma(1 - b_i)}}{(n!)^{q-1}}
\,\,  ,
\end{equation}
which gives for $\NN(t)$ exactly the generalized hypergeometric function, that is,
\begin{equation}
\label{Ngen }
\NN(t) = \,  _qF_q( \, \underset{q-times}{\underbrace{1, 1, ...,1}};
1 - b_1,1 -  b_2, ...1 - b_q; t)\,\, .
\end{equation}
When $q = 1$ and $b_1 = -a$, we have
\begin{equation}
\label{q1 }
x_n = n + a \, ,  \quad n \geq 1\, , \ x_0 = 0\, ,
\end{equation}
and
\begin{equation}
\label{Nq1 }
\NN(t) = \,  _1F_1(1; 1 + a; t)\,\, ;
\end{equation}
for $q = 2$,
\begin{equation}
\label{q2 }
x_n = \frac{(n - b_1) (n - b_2)}{n} \, ,  \quad n \geq 1\, , \ x_0 = 0\, , 
\end{equation}
and
\begin{equation}
\label{Nq2 }
\NN(t) = \,  _2F_2(1,1; 1 - b_1, 1 - b_2; t)\,\, .
\end{equation}

The example (\ref{xisn2}) solved above is a $q=2$ case, with
$a_1 = \sqrt{\epsilon}$ and $a_2 = -\sqrt{\epsilon}$.

There are sequences still more general than (\ref{alphagen}) that result in
perturbations of the exponential $\NN(t)$, that is,
\begin{equation}
\label{alphamgen }
\alpha^{(q)}(n) = n \frac{\prod_{i=1}^q(n - b_i) -
\prod_{i=1}^q(n - a_i)}{\prod_{i=1}^q(n - a_i)}\,\,.
\end{equation}
The Delone sequences for appropriate values of the constants $\{ a_i \}$
and $\{ b_i \}$ are then
\begin{equation}
\label{xismgen}
x_n = n \frac{\prod_{i=1}^q(n - b_i)}{\prod_{i=1}^q(n - a_i)} \, ,  \quad n \geq 1\, , \ x_0 = 0\, , 
\end{equation}
and our ``perturbed exponential" becomes
\begin{equation}
\label{fmoregen }
\NN(t) = \, _qF_q(1 - a_1, 1 - a_2, ...,1 - a_q; 1 - b_1,1 - b_2, ..., 1 - b_q; t)
\,\,  .
\end{equation}

The Delone sequence (\ref{conf1}) is a particular case of (\ref{xismgen})
with $q = 1$, $a_1 = 1 - a$ and $b_1 = 1 - a - \epsilon$. The constants $\{ a_i \}$
and $\{ b_i \}$ have to be chosen in such a way that the
values of $\alpha^{(q)} (n)$ satisfy the criteria required by the Delone
sequences, for any value of $n$.


\subsection{Asymptotic estimates}

Let us now analyze the asymptotic behavior of some sequences presented in the
last paragrah for very large values of $n$.

As a first example, let us take the Delone sequence (\ref{conf1}) for the
special value $a = 1$, i.e. the constant shift for $n \geq 1$.  In (\ref{mun1}), note that the numerator is the Gamma
distribution which is centered around a value of $t$ of the order of $n$. When
we consider
the limit for very large $n$, only large values of $t$ contribute to the
integral. But, for these values of $t$,
\begin{equation}
\label{small }
\frac{\epsilon \, \Gamma(\epsilon, t)}{\Gamma(\epsilon + 1)} << 1\,\,\,   ,
\end{equation}
and we have for $\mu_n$ the approximate expression 
\begin{equation}
\label{muasy}
\mu_n \cong 1 + \frac{\epsilon\,\, \Gamma(n + 2\epsilon + 1)}{
\Gamma(\epsilon + 1)\, \Gamma(n + \epsilon + 2)\, 2^{n+2\epsilon}} \, F(1, 1 -
\epsilon; n + \epsilon + 2; -1) \,\, .
\end{equation}

\begin{figure}[h]
\begin{center}
\includegraphics[width=5.5in]{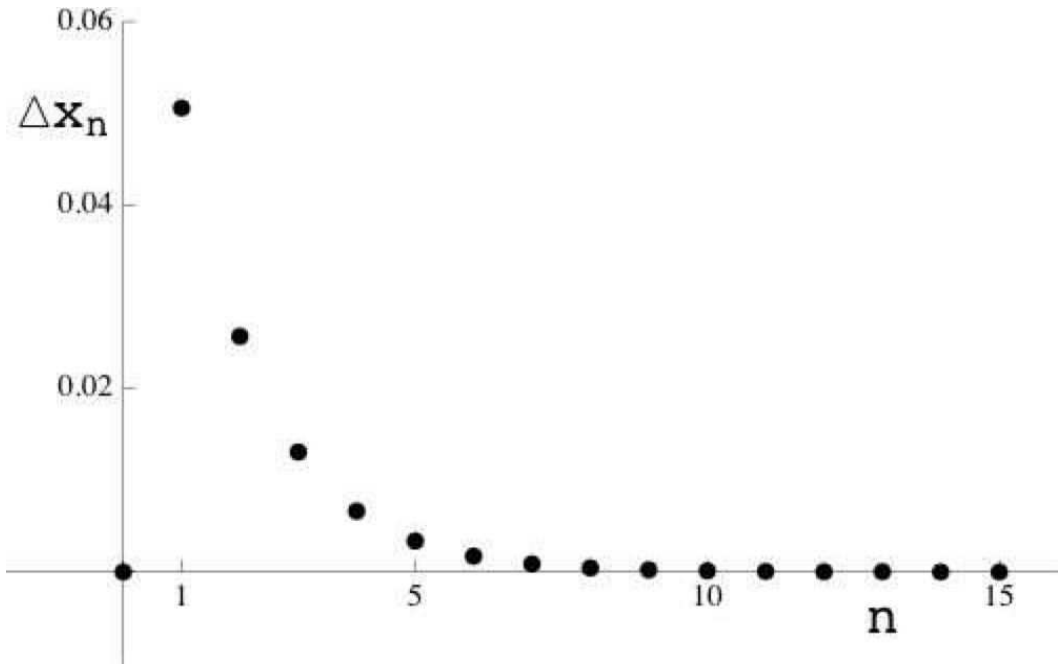}
\caption{Delone sequence  $x_n = n + \epsilon\, \, ,   n \geq 1$, and  $x_0 = 0$.
Behavior of
$x_n - \tilde{x}_n $ for the first 15 values of $n \geq 0$ ($\epsilon = 0.1$).}
\label{constant}
\end{center}
\end{figure}

For large n, the hypergeometric function $F(1, 1 - \epsilon; n + \epsilon + 2; -1)$
becomes a finite series; as $n$ is much larger than $\epsilon$, we can keep only
its first term. Therefore, we finally obtain that the asymptotic expression for
$\mu_n$ is
\begin{equation}
\label{mufinal}
 \mu_n\,\,\, \underset{n \rightarrow  \infty}{\longrightarrow } \,\, 1 +
 \frac{1}{\Gamma(\epsilon) \,
 n \,  2^n }\,\, ,
\end{equation}
which is pratically $1$ for $n= 10$.
In order to have a better idea of the perturbation in the spectrum $\tilde{x}_n$
introduced
by $\mu_n$, (see equation  (\ref{seqren})), we show in Figure \ref{constant}
the behavior of
$x_n - \tilde{x}_n $ for the first 15 values of $n$ ($\epsilon = 0.1$).

As a second example, let us consider the particular case of (\ref{homseq})
for which $a = d = 0$ and $b/c = -\epsilon$, where $\epsilon$ is a small
real parameter. In this case, $x_n$ is given by the equation (\ref{xisn2}),
which leads to $x_n!$ and $\NN(t)$ shown by Equations (\ref{xisn2fat})
and (\ref{2f2}). $\mu_n$ can
then be written as
\begin{equation}
\label{munn2}
\mu_n = \frac{1}{x_n!} \int_0^\infty \frac{t^n dt}{_2F_2 (1,1; 1-\sqrt{\epsilon},
1+\sqrt{\epsilon};t)} \, .
\end{equation}
For large values of $n$, the main contribution for the integral of the previous
equation is
given by $t \sim n$ and the dominant contribution of the hypergeometric for $t>>1$ is
\begin{equation}
\label{dom2f2}
 _2F_2(1,1; 1-\sqrt{\epsilon},1+\sqrt{\epsilon};t) \underset{t>>1}{\sim}
\Gamma \left[1-\sqrt{\epsilon}\right] \Gamma
\left[1+ \sqrt{\epsilon} \right] e^t \left(1  -\frac{\epsilon}{t} - \frac{(1-
\epsilon) \epsilon}{2 t^2}\right) \, ,
\end{equation}
leading to the following asymptotic behavior for $\mu_n$:
\begin{equation}
\label{asinmun2}
  \mu_n    \underset{n \rightarrow  \infty}{\longrightarrow }1 +
  \frac{\epsilon}{n^2}  \, ,
\end{equation}
that presents the same behavior of $x_n/n$ in function of $n$ as shown
in Equation (\ref{xisn2}).

\begin{figure}[h]
\begin{center}
\includegraphics[width=5in]{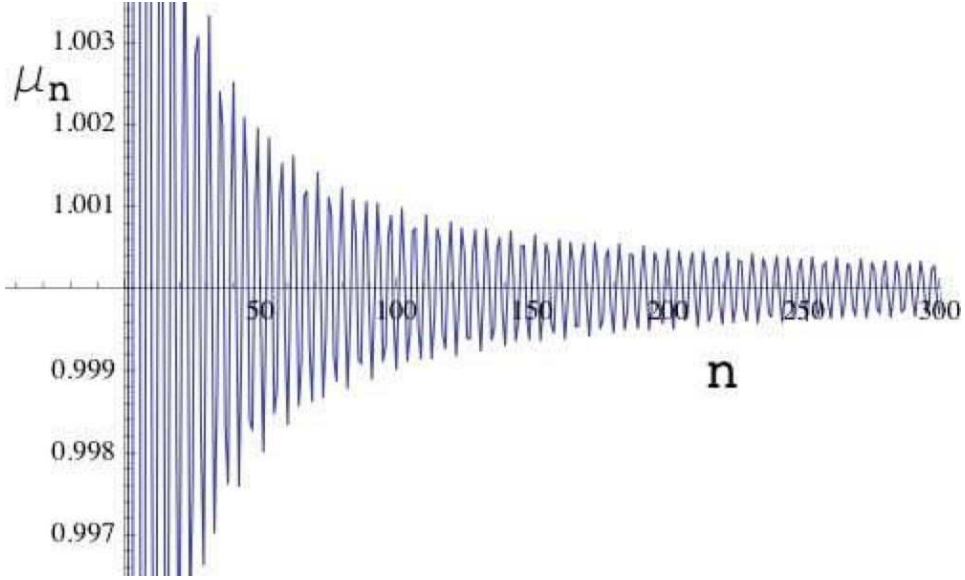}
\caption{Delone sequence $x_n = n + 0.1\, \dfrac{\sin{\sqrt{2}n}}{n}\, ,   \quad n \geq 1\, , \ x_0 = 0$. Behavior of $\mu_n$,
for the first 300 values of $n \geq 1$. The large oscillations present near the origin have been truncated for convenience.}
\label{figsin}
\end{center}
\end{figure}

A more complicated example of a Delone sequence with a decay in $n$ similar to the
first example of this subsection but presenting an oscillating character,
is the sequence given by the following expression for $\alpha(n)$:
\begin{equation}
\label{alfaseno}
\alpha (n) = \frac{\epsilon \sin(\sqrt{2} \, n)}{n} \, ,
\end{equation}
where $\epsilon$ is a small parameter. In this case, $x_n!$, $\NN(t)$
and $\mu_n$ can be calculated only numerically.  We have  found that  the sequence
$\mu_n$ oscillates, and  the asymptotic behavior of its maxima is
$ | \mu_n -1 |  \equiv \epsilon / n$. In Figure \ref{figsin} we show $\mu_n$
for the first $300$ values of $n$, taking $\epsilon = 0.1$ in (\ref{alfaseno}).


\section{Sequences and triplets: a general algebraic setting}

\subsection{Heisenberg-like triplet}

Suppose we are given a finite or infinite,  sequence of
nonnegative real numbers, bounded from below,
$\{ x_n \}_{n\in \N}$ such that $x_0 = 0$. Let us view this sequence as
the set of eigenvalues for a
positive self-adjoint operator denoted by $x_N$.
\be
x_N e_n = x_n e_n.
\en
The associated separable Hilbert space ${\cal H}$ is precisely the closure
of the linear span of
the orthonormal system $\{ e_n \}$, {\it i.e.} the latter is an
orthonormal basis for ${\cal H}$. We
now define the operator  labeling the states, or number operator, by $N$,
\be
N e_n = n e_n,
\en
providing also the rationale for the notation  $x_N$, or more
generally of the notation $f(N)$ for any diagonal
real operator defined by $f(N) e_n = f(n) e_n$ constructed from a sequence
$n \rightarrow f(n)$. The set
$\R^{\N}$ of these sequences is a commutative algebra on $\R $, and this
naturally induces a
structure of a real commutative algebra for the corresponding set of
$f(N)$. This algebra will be
denoted by ${\cal D}$.  We now define lowering and raising operators for
the sequence  $\{ x(n) \} $:
\be
a \, e_n =\sqrt{x_n} e_{n-1}\, , \quad a^{\dagger} \, e_n =  \sqrt{x_{n+1}}
e_{n+1}\, .
\en
From these definitions it is clear that
\be
\label{xN}
x_N = \ad a.
\en
Now, for any diagonal
operator $f(N)$ in ${\cal D}$, the following commutation rules hold:
\be
a f(N) = f(N+1) a \equiv f_{(-1)}(N)a, \ f(N)a^{\dagger} = a^{\dagger}
f_{(-1)}(N),
\en
\be
\lbrack a, f(N) \rbrack = f'(N)a = a f_{(1)}'(N), \ \lbrack a^{\dagger},
f(N) \rbrack = - a^{\dagger}
f'(N) = -f_{(1)}'(N) a^{\dagger},
\en
where $f_{(m)}(N) \equiv f(N-m), \, m\in \mathbb{Z}$, is the $m^{\mbox{th}}$
shift of $f$, with
the convention that $f(n)=0$ for $n < 0$, and  $f'(N) \equiv f(N+1) -
f(N)$ is the finite
difference
derivative of $f(N)$. Of course, the validity  (or otherwise) of the above
relations depend on domain considerations for the operators in
question. In particular, we also have
\be
\lbrack a, a^{\dagger} \rbrack = x'_N.
\en
The triplet (lowering $a$, raising $a^{\dag}$, commutator $\lbrack a,
a^{\dag} \rbrack = x'_N
$) will be called a spectrum generating triplet of the sequence  $\{ x_n \}
$. In \cite{galawin} the Lie algebra  generated by
such a triplet  has been  investigated in a systematic way .

\subsection{Generalized Heisenberg triplet}

An algebraic structure which contains $q$-oscillators as a particular case was
proposed in \cite{jpa} and has been successfully used in some different physical
problems \cite{nmon,jnc,bcr,ricure,hcrl,hcrl2}.
In this new algebra, called Generalized  Heisenberg Algebra (GHA),
the commutation relations among the operators $a$, $a^{\dagger}$ and a diagonal operator $J_0$ which is going to replace in the present context  $x_N$ introduced in (\ref{xN}) depend on
a general functional $h(x)$ and  are given by:
\begin{equation}\label{eq0a}
J_0a^\dagger =a^\dagger h(J_0)
\end{equation}
\begin{equation}\label{eq0b}
aJ_0=h(J_0)a
\end{equation}
\begin{equation}\label{eq0c}
[a ,a^\dagger ]=h(J_0)-J_0,
\end{equation}
with  $a=(a^{\dagger})^{\dagger}$ and $J_0=J_{0}^{\dag}$.
This algebra tells us that
the $J_0$ eigenvalues ($J_0 \, e_n=t_n \, e_n$)
are obtained by a
one-step recurrence relation ($t_n=h(t_{n-1})$), {\it i.e},
each eigenvalue depends on the
previous one.  Thus, the eigenvalue behavior can be studied by dynamical system
techniques, simplifying the task of finding possible representations of the
algebra \cite{jpa}. In order to relate this algebra to a physical system, the
$J_0$ operator can be identified with the Hamiltonian of the system.

When the functional $h(x)$ is linear we recover  \cite{jpa} the $q$-oscillator
algebra; other types of functionals lead to other different
algebraic structures \cite{hugo}. The functional $h(x)$ can be, for
example, a polynomial and hence depend on some parameters
(the polynomial coefficients).
Let us consider the vector $e_0 $, with the lowest eigenvalue of $J_0$,
$J_0 \,e_0 = t_0 \, e_0$.
For a general $h(x)$, these operators act on the Fock space as
\begin{eqnarray}
J_{0} \,e_m &=& h^{(m)}(t_0) \, e_m , \; \; \; m = 0,1,2,
\cdots \; ,
\label{eq:b1} \\
a^{\dagger} \, e_m &=& N_{m} \, e_{m + 1} ,
\label{eq:b2} \\
a \,e_m &=& N_{m-1} \, e_{m - 1},
\label{eq:b3}
\end{eqnarray}
where $N_{m-1}^2 = h^{(m)}(t_0)-t_0$ and $ h^{(m)}(t_0)$
is the $m$-th iteration of $t_0$ by means of the function $h$.

It is interesting to observe that in some cases we can find a GHA associated
to Delone perturbations of $\mathbb{N}$. Indeed, let us  consider those
Delone sequences, eq. (\ref{delpert}), that can be inverted. In those cases
we can write
\begin{equation}
\label{delpertinv}
n = x_n + \gamma(x_n) \, .
\end{equation}
Note that the difference $(x_n - n)$ implies that $\gamma(x_n)$ shares  the same
properties as $\alpha(n)$. Therefore, the inverse sequence is also a Delone
sequence.

In order to find the GHA for invertible Delone sequences, we write the
$(n+1)$-st term of the  sequence in terms of $x_n$:
\begin{equation}
\label{delpertinv2}
x_{n+1} = x_n + 1 + \delta(x_n) \, ,
\end{equation}
where
\begin{equation}
\label{xxx}
\delta(x_n) = \gamma(x_n) + \alpha(x_{n}+1 + \gamma(x_{n})  ) \, .
\end{equation}
The characteristic function is trivially obtained from eq. (\ref{xxx}):
\begin{equation}
\label{efe}
h(x) = x + 1 + \delta(x) \, ,
\end{equation}
and the corresponding GHA is  then
\begin{equation}\label{eq00a}
[J_0, A^\dagger] =A^\dagger (1+ \delta(J_0))
\end{equation}
\begin{equation}\label{eq00c}
[a,a^\dagger ]= 1 + \delta(J_0) \, .
\end{equation}
From eqs. (\ref{delpert}) and (\ref{delpertinv}) it can be seen that $\alpha$ and
$\gamma$ are perturbations of opposite signs.  Thus  $\delta$ is the difference
of two perturbation terms and hence itself a perturbation. Eqs. (\ref{eq00a}) and
(\ref{eq00c}) show that the invertible Delone sequences  associated to GHA are
perturbations of the quantum harmonic oscillator. 
\begin{figure}
\begin{center}
\includegraphics[width=4in]{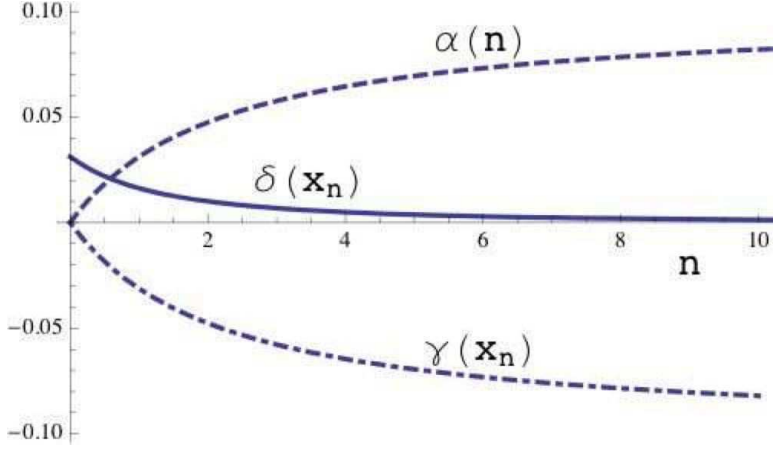}
\caption{Delone sequence $x_n = n + \alpha(n)$ with $\alpha(n) = \frac{a n }{c n + d}$, for the case $a = 0.1$, $c=1$, and $d=2.2$. The behavior of $\alpha(n)$
(dashed line), $\delta(x_n)$ (continuous line) and $\gamma(x_n)$
(dot-dashed line) is shown for the first 10 values of $n$.}
\label{figAlg1}
\end{center}
\end{figure}

We take as an example of an  invertible Delone sequence the case
\begin{equation}
\label{proj}
\alpha(n) = \frac{a n }{c n + d}\,, \quad n \geq 0\,,  \quad d \neq 0\, , 
\end{equation}
where the real constants $a$, $c$ and $d$ have to satisfy the following conditions:
\begin{equation}
\label{conditions}
\left \vert \frac{a}{c} \right \vert  < 1 ; \,\,\,\,\,\,c \neq 0;  \,\,\,\,\, -\frac{4}{c^2} \, ad \geq r-1\, , 
\end{equation}
and $r\in (0,1)$, according to Definition 2.1.

 For this sequence,
\begin{equation}
\label{gamma1}
\gamma(x_n) = \frac{\sqrt{c^2 x_n^2 + 2 c (d-a) x_n + (a+d)^2 } -
c x_n - a -d}{2 c}
\end{equation}
and
\begin{equation}
\label{delta1}
\delta(x) = \gamma(x) + \frac{a(x + \gamma(x)) + a }{c (x + \gamma(x)) + c + d} \, .
\end{equation}


For the particular case
$x_n = n + \alpha(n)$ with $\alpha(n) = \frac{a n }{c n + d}$, see Figure \ref{figAlg1} where the functions  $\alpha$, $\gamma$
and $\delta$ are explicitly exhibited.

 If we choose specifically $a= \epsilon$, $| \epsilon| <1$,    $c=1$ and $d=0$, then  $\alpha(n) =
  \epsilon$, 
  $\delta  =0$  and the GHA is given by $h(x) = x+1$ which would characterize  the harmonic oscillator algebra. However this sequence is out the scope of the present paper since $x_0 = \epsilon \neq 0$. Imposing $x_0 = 0$ besides $x_n = n + \epsilon$ for $n \geq 1$ takes us to the case already examined in the previous section and for which we do not have a GHA algebraic structure. 
   

\section{A result on asymptotic Poisson and Gamma distributions}\label{poissgam}
Let us now present a  result which explains  the behavior of the sequence
$(\mu_n)$ given a certain class of Delone sequences, namely the existence of
$\mu_{\infty}$ for  $\alpha$-Delone  perturbations of $\N$.
For  an $\alpha$-Delone perturbation $x_n = n + \alpha(n)$ of $\N$
one can write
\begin{equation}
\label{factorxi}
x_n! = n! \, \frac{1}{\xi(n)}\quad \mbox{with} \quad \frac{1}{\xi(n)} \deq
\prod_{k=1}^n\left(1 + \frac{\alpha(k)}{k}\right)\, .
\end{equation}
 Then the associated ``exponential'' reads as
 \begin{equation*}
\mathcal{N}(t) = \sum_{n = 0}^{\infty} \frac{t^n}{n!}\, \xi(n)\, ,
\end{equation*}
Its ratio to the ordinary exponential reads as the Poisson average of the
random variable $n \mapsto \xi(n)$
\begin{equation*}
\mathcal{N}(t) \, e^{-t} =  \sum_{n = 0}^{\infty} \frac{t^n}{n!}\, e^{-t}
\xi(n) \deq (\mathrm{E}_{P}\xi)(t) \, ,
\end{equation*}
Thus, the ratio $\mu_n/\xi(n)$ can be rewritten as the  Gamma average of
the random variable $t \mapsto e^t/\mathcal{N}(t)$ :
 \begin{equation*}
\label{gamv}
\frac{\mu_n}{\xi(n)} = \frac{1}{\xi(n)} \int_{0}^{+\infty}
\frac{dt}{\mathcal{N}(t)}\, \frac{t^n}{x_n!} = \int_{ 0}^{\infty}
\frac{t^n}{n!}\, e^{-t}\, \frac{e^t}{\mathcal{N}(t)}\, dt \deq
\left(\mathrm{E}_{G} \frac{1}{(\mathrm{E}_{P}\xi)(t)}\right)(n)\, .
\end{equation*}

Let us put this on a more general setting.
Let $\N \ni n \mapsto \xi(n)$
be a discrete function which is
extendable to a function $\R^+ \ni x \mapsto \R
$ with Ê$\xi(0) = 1$ and $\lim_{x \to \infty} \xi(x) = 0.$
Its Poisson mean value with parameter $t$ is given by
\begin{equation}
\label{poissmv}
(\mathrm{E}_{P}\xi)(t) := \sum_{n = 0}^{\infty} \frac{t^n}{n!}\, e^{-t} \xi(n)\, ,
\end{equation}
whereas the Gamma mean value with parameter $m$ of a random variable $\Xi$ is
given by
\begin{equation}
\label{gamv1}
(\mathrm{E}_{G}\Xi)(m) = \int_{ 0}^{\infty} \frac{t^m}{m!}\, e^{-t}
\Xi(t)\, dt\, .
\end{equation}
Let us examine the asymptotic behavior of the following combination of these
two averages:
\begin{equation}
\label{intIM}
\mathrm{I}_m =\xi(m) \, \left( \mathrm{E}_{G}\frac{1}{(\mathrm{E}_{P}\xi)}
\right)(m) = \int_{0}^{\infty} \dfrac{\dfrac{t^m}{m!}\, \xi(m)}{\sum_{n =
ÊÊ0}^{\infty} \dfrac{t^n}{n!}\, Ê\xi(n)}\, dt\, .
\end{equation}
We would like to give sufficient conditions for having $\lim_{m \to \infty}
\mathrm{I}_m = 1$.

\begin{Prop}
\label{propphi}
Let us define the logarithm of $1/\xi$ by $\phi$ :
\begin{equation}
\label{xiphi}
\xi(x) = e^{-\phi(x)}\, ,
\end{equation}
and suppose that the Êfunction $\R^+ \ni x \mapsto \phi(x)$ has the following properties.
\begin{itemize}
\item[(i)] $\phi(x)$ is three-times derivable almost everywhere (a.e.) on $\R^+$,
\item[(ii)] $\vert \phi'(x)\vert \ll 1 $(a.e.) at large $x$.
Also $x \vert \phi''(x)\vert \ll 1 $(a.e.) and
$x^2 \vert \phi'''(x)\vert \ll 1 $(a.e.) at large $x$.
\end{itemize}
Then $\mathrm{I}_m \underset{ m \to \infty}{\to } 1\, .$
\end{Prop}

\DEM \\
Let us put
\[ A(t)\equiv \sum_{n = 0}^{\infty} \frac{t^n}{n!}\, Ê\xi(n)
=\frac{1}{\sqrt{2 \pi}}
\sum_{n = 0}^{\infty}e^{F_t(n)},
\]
with $\xi(x)\equiv e^{-\phi(x)}$ and
\begin{equation}
\label{sumint2}
F_t(x) := x\ln{\frac{t}{x}} + x - \frac{1}{2}\ln{x} - \phi(x)
\end{equation}
where we have neglected the errors in the Stirling's formula, $n^n
e^{-n}(\sqrt{2\pi n})/n!=1+{\cal O}(n^{-1})$ (e.g.
${\cal O}(n^{-1})=-0.012$ for $n=7$.)
It then follows
\[ F_t'(x) = \ln{\frac{t}{x}} - \frac{1}{2x} - \phi'(x)\,, Ê\quad F_t''(x) =
-\frac{1}{x} + \frac{1}{2x^2} - \phi''(x)\, .\]
We assume that there exists a  solution to
$ F_t'(x) = 0 $ at $x=x^{\ast}(t)$. This implies
\begin{equation}
\label{xaVSt}
\xa e^{\frac{1}{2\xa}+\phi'(\xa)}= t Ê.
\end{equation}
We now impose two conditions.
\vskip 0.2cm
\noindent
{\bf Requirement 1:}\footnote{This requirement comes from the application of the
saddle point method.}\emph{We demand that  
[$t\to \infty$] implies [$\xa(t)$ exists \underline{and} that $\xa(t)\to\infty$}]\\

In order for $\xa(t)$ to exist  at large $t$,
the function $x e^{\frac{1}{2x}+\phi'(x)}$ needs to be unbounded. We face  two cases:
\begin{itemize}
  \item (Case-a) If it diverges for finite $x$ due to a singular behavior of $\phi'(x)$,
then the second condition ($\xa(t)\to\infty$) is violated.
Therefore this case is excluded,
i.e. $\phi'(x)$ should be locally bounded. 
  \item (Case-b) The asymptotic behavior of $\phi(x)$ for $x\gg 1$
is such that it allows the divergence of $x e^{\frac{1}{2x}+\phi'(x)}$
in this limit, i.e.\footnote{$x e^{\frac{1}{2x}}\simeq(x+\frac{1}{2})$
for $x\gg 1$.} $\phi'(x)$ can be negatively large but it does not dominate
$\ln x$. That is, either [$\phi'(x)>0$ ] or
[$\phi'(x)<0$ and $|\phi'(x)|\ll\ln x$] for $x\gg 1$. \end{itemize}
\noindent
In summary
\[ \phi'(x) \ge -\ln x +\mbox{(positive increasing)}
\quad (\mbox{for }x\to\infty)\]
The condition for $\phi(x)$ in the hypothesis is modest enough to satisfy the
above condition.

\vskip 0.2cm
\noindent
{\bf Requirement 2:}\footnote{This requirement also comes from the application of the
saddle point method.} \emph{For $\xa\gg 1$, the second derivative at the peak
$\vert F_t''(\xa)\vert$ is negative, and
\[ \vert F_t''(\xa)\vert^{-\frac{1}{2}} \ll \xa \]
so that the hump of the exponential becomes asymptotically sharp.}
\noindent
The above inequality is equivalent to (assuming that the negative value condition
is satisfied)\footnote{We can neglect $-\frac{1}{2x^2}$ with respect to
$\frac{1}{x}$}
\[\frac{1}{x}+\phi''(x) \gg \frac{1}{x^2}.\]
The second derivative $\phi''(x)$ can be negative, but does not dominate or cancel the $\frac{1}{x}$.
We formulate this condition as
\[\phi''(x)\ge -\frac{1}{x}+ \eta(x), \quad \eta(x)\gg \frac{1}{x^2}
\quad (\mbox{for }x\to\infty)\]
The condition for $\phi(x)$ in the hypothesis is modest enough to satisfy the
above condition.\\

\vskip 0.2cm
\noindent
{\bf Application of  the Laplace method:}
\vskip 0.2cm
With the above two requirements being satisfied, the evaluation of
the sum is done by using
\[e^{ F_t(x)}\simeq e^{ F_t(\xa)}
e^{-\dfrac{1}{2}\, \vert F_t''(\xa)\vert \, (x-\xa)^2}. \]
The result is \footnote{We used $
\xa\, \vert F_t''(\xa)\vert\simeq 1+\xa \phi''(\xa)$ for
$\xa\gg 1$.}
\[
A(t)\simeq
\frac{\mathrm{exp}{ (\frac{1}{2} + \xa\, \phi'(\xa) + \xa - \phi(\xa)
)}}{\sqrt{ 1+\xa \phi''(\xa) }}\, \]
Note that $\frac{1}{\sqrt{ 1+\xa \phi''(\xa) }}\ll \sqrt{\xa}.$

Next we evaluate

\[ \label{eq:Im}
\mathrm{I}_m =\int_0^\infty \frac{t^m \xi(m)}{m! A(t)}dt
\simeq
\frac{1}{\sqrt{2\pi m}}\int_0^\infty e^{G(t)}dt,
\]%
with
\begin{eqnarray}
G(t)&=& m\ln\frac{t}{m}+m-\phi(m)-\ln A(t) \cr
&\simeq &
m\ln\frac{t}{m}+(m-\xa)-[\phi(m)-\phi(\xa)] \cr && -\frac{1}{2}-\xa \phi'(\xa)+
\frac{1}{2}\ln (1+\xa \phi''(\xa))
\end{eqnarray}
The time derivative of the defining equation of $\xa$
(eq.\ref{xaVSt}) gives
\[\frac{d\xa}{dt}=\frac{e^{\frac{1}{2\xa}+\phi'(\xa)}}{1+\xa \phi''(\xa)}
\quad (\mbox{for }t\to\infty)\]

Then
\begin{eqnarray}
G'(t)&=&\frac{m}{t}+\frac{d\xa}{dt}\left[
-(1+\xa \phi''(\xa))+\frac{\phi''(\xa)+\xa \phi'''(\xa)}{2(1+\xa \phi''(\xa))}
\right]\cr
&=&\frac{m}{t}+
e^{\frac{1}{2\xa}+\phi'(\xa)}
\left[-1+\frac{\phi''(\xa)+\xa \phi'''(\xa)}{2(1+\xa \phi''(\xa))^2}
\right]
\end{eqnarray}
Note that the denominator in [\,] is such that
$(1+\xa \phi''(\xa))^{-2}\ll \xa$.\\
\vskip 0.2cm
\noindent
{\bf Final concession:}
\vskip 0.2cm
{\it Here we shall introduce further simplifying assumptions (at the
possible cost of narrowing the validity region of $\phi(x)$)\footnote{
If we wish to improve the validity of the theorem, we should put
more precise conditions here.}:}
\[ \xa \phi''(\xa) +{\xa}^2 \phi'''(\xa)\to 0
\quad (\mbox{for }t\to\infty)\]
\[\frac{1}{2\xa}+\phi'(\xa)\to 0
\quad (\mbox{for }t\to\infty)\]
These conditions are just fulfilled under the hypotheses of the proposition.\\
\vskip 0.2cm
\noindent
{\bf Second application of the  Laplace method:}
\vskip 0.2cm
Then we have first $\xa \simeq t$ asymptotically, and
$G'(t)\simeq \frac{m}{t}-1.$ Therefore,
\[G'(m)=0 \quad (\mbox{for }t\to\infty)\]
and $G''(t)=-\frac{m}{t^2}$ gives
\[G''(m)=-\frac{1}{m}.\]
The Laplace method then yields
\[ \mathrm{I}_m\simeq
\frac{1}{\sqrt{2\pi m}}\int_0^\infty e^{G(t)}dt
\simeq
\frac{1}{\sqrt{2\pi m}}\int_{-\infty}^\infty e^{\frac{(t-m)^2}{2m}}dt =1.
\]

\vskip 0.2cm
\noindent
{\bf Implications :}
\vskip 0.2cm
As for $\xi(m)=e^{-\phi(m)}$, the conditions
\[\phi'(x)\to 0,\quad Êx\phi''(x)\to 0, \quad Êx^2 \phi'''(x)\to 0
\quad (\mbox{for }t\to\infty)\]
are rather loose.
$\phi(x)=(-a) \ln x$ with both $a>0$ and $a<0$ are permissible.
That is $\xi(m)$ can grow or decrease as a power law, $\xi(m)\sim x^{a}.$
So $\phi(x)=a x^q$ with $q<1$ and both $a>0$ and $a<0$ are  admissible.
That is, $\xi(m)$ can even grow like $e^{a m^q}$ with $a>0$ and $q<1$.

\section{Application to Delone $\alpha$-perturbations of $\N$}

Let us consider an $\alpha$-perturbation of $\N$
\begin{equation}
\label{delpert1}
x_n = n + \alpha(n)\, , \quad n \in \N\, ,
\end{equation}
where $ \N \ni n \mapsto \alpha(n) $ is
 a bounded function with values in the interval $(-1,1)$, $\alpha(0) = 0$,
 and such that its successive jumps $\alpha(n+1) -\alpha(n)$ have lower
 bound $r-1$ with $r\in (0,1)$.

One can write $x_n! = n! \, \frac{1}{\xi(n)}$ with
\begin{equation*}
\frac{1}{\xi(n)} = \prod_{k=1}^n\left(1 + \frac{\alpha(k)}{k}\right)\, .
\end{equation*}

Then the logarithm of $1/\xi$ reads as :
\begin{equation}
\label{xiphi1}
\phi (n) = \sum_{k=1}^n \ln{\left(1 + \frac{\alpha(k)}{k}\right)}\, .
\end{equation}

Let us extend the discrete domain of this  function to a continuous one
by replacing   the sum by the integral

\begin{equation}
\label{contphi}
\phi(x)  = \int_1^x \ln{\left(1 + \frac{\alpha(y)}{y}\right)} \, dy\, .
\end{equation}

Hence,
\begin{equation}
\label{derphi1}
\phi'(x)= \ln{\left(1 + \frac{\alpha(x)}{x}\right)} \,
\underset{\mbox{at large $x$}}{\approx} \frac{\alpha(x)}{x}
\end{equation}
and so $\phi'(x) \to_{x \to \infty} 0$ since $x \mapsto \alpha(x)$ is bounded.
For the second derivative, we have
\begin{equation}
\label{derphi2}
\phi''(x)= \frac{x\alpha'(x) - \alpha(x) }{x^2 + x \alpha(x)} \,
\underset{\mbox{at large $x$}}{\approx} \frac{\alpha'(x)}{x}\, .
\end{equation}

Therefore, from Proposition \ref{propphi} one can assert the following.
\begin{Prop}
\label{delphi}
Let  $x_n = n + \alpha(n)\, ,\, n \in \N\,,$ an $\alpha$-perturbation of $\N$
such that the map $\N \ni n \mapsto \alpha(n)$ extends to a function
$\R^+ \ni x \mapsto \alpha(x)$ that is a.e. two times derivable. Suppose that
$\vert\alpha'(x)\vert \ll 1$ and $x\vert\alpha''(x)\vert \ll 1$ at large $x$. Then we have
\begin{equation}
\label{asympdel}
\mathrm{I}_m  = \int_{0}^{\infty} \dfrac{\dfrac{t^m}{x_m!}}{\sum_{n = 0}^{\infty}
\dfrac{t^n}{x_n!}}\, dt\, \to_{m \to \infty} \, 1\, .
\end{equation}
 \end{Prop}

An example of such a sequence is  given by the periodic function :

\begin{equation}
\label{fibex}
\R_{\ast}^+ \ni x \mapsto \alpha(x) = \lambda\{\mu x \} + \nu\, ,
\end{equation}
with $\vert \lambda \vert < \frac{r-1}{2}\, , \, r\in (0,1)$,
$\vert\alpha'(x)\vert = \vert \lambda \mu\vert $ a.e.,  $\alpha''(x) = 0$ a.e., and
\begin{equation*}
\begin{array}{ccc}
 -1 <  \nu  <1 - \lambda  &\mbox{if} & \lambda > 0 \, ,    \\
     -\lambda  -1 <  \nu < 1   &\mbox{if} & \lambda < 0   \, .
\end{array}
\end{equation*}
Note that  the sequences $\Z_{\beta}^+/c_{\beta}$ of rescaled non-negative $\beta$-integers with $\beta$ a quadratic PV unit, such they are defined in Proposition \ref{unit_gago}, are of this type and so fulfill the conditions of Proposition \ref{delphi}.
\section{Discussion and outlook}

In this work we have explored  the possibility of quantizing the
complex plane through the use of sequences of numbers close
enough to the natural numbers. We have discussed the probabilistic
content of the procedure. Some algebraic aspects have also been
considered. We have treated a specific example, the hypergeometric
case, which leads to explicit estimates. A general mathematical
result has also been presented which pertains to the present
context as well as to mathematical statistics.  Many aspects of our work deserve further investigation,
on different levels,  numerical, mathematical, and interpretational,
specially those around the notion of quantum or fuzzy  localization
in the  complex plane arising from such a ``non-commutative'' reading
of the complex plane.

Let us just discuss, from a more oriented physical point of view, the  relations between the original sequence $\mathcal{X} = \{ x_n\, , \, n \in \N\}$, its renormalized companion $\widetilde{ \mathcal{X}} = \{\tilde x_n\, , \, n \in \N\}$ and the modified Lebesgue measure on the complex plane, $ \nu(dz) = \dfrac{\widetilde{\mathcal{N}}(\vert z \vert^2)}{\mathcal{N}(\vert z
\vert^2)}\, \dfrac{d^2 z}{\mu_0\, \pi}$. Suppose we observe through some experimental device the sequence of numbers $\mathcal{X} $, for instance the  quantum energy spectrum of a given system,  that lies in the class of $\alpha$-perturbation of the nonnegative integers. Due to this proximity, probably the considered system is  classically described by  a harmonic oscillator with Hamiltonian $H = \frac{1}{2} (p^2 + q^2)$. Since the canonical quantization of the harmonic oscillator yields the integers as a spectrum, we here attempted to set up a quantization framework which takes account of the modified spectrum. For this purpose, we need to solve a Stieltjes moment problem. We know that this is impossible  for most of  sequences $\mathcal{X} $, for instance the sequences of rescaled beta-integers mentioned  in this paper. Hence we are naturally led to deal with the renormalized  sequence $\widetilde{ \mathcal{X}}$ for which we solve the moment problem with the measure $\dfrac{\widetilde{\mathcal{N}}(t)}{\mathcal{N}(t)}\, \dfrac{dt}{\mu_0}$. This measure endows the complex plane viewed as the phase space for the harmonic oscillator with a statistical content, in the sense that classical states are not anymore described by points $z_0$, i. e. by Dirac distributions $z \mapsto \delta_{z_0} (z)$, but instead by a smooth distribution of the type  $ z \mapsto\dfrac{\widetilde{\mathcal{N}}(\vert z -z_0 \vert^2)}{\mathcal{N}(\vert z -z_0\vert^2)}$. Then, by following the
CS quantization stemming from the renormalized sequence, we consistently find the latter as the renormalized version of the observed spectrum along the equation
\begin{equation*}
\label{qcsho}
\widetilde{A}_{\frac{1}{2} (p^2 + q^2)}= \widetilde{A}_{z\bar z} = \tilde{a}\, \tilde{a}^{\dag} =\tilde x_{N+1}\, .
\end{equation*}
This spectrum is not the observed one, but it might differ appreciably of the later in the first levels only, since $\tilde x_n = (\mu_n/\mu_{n-1})\, x_n$. Actually only the differences $\tilde x_{n+ m} -   \tilde x_n$ are experimentally significant  and we should reasonably expect that they are  practically the same as  $ x_{n+ m} -    x_n$ for excited levels, as it can be observed in the particular example shown in Figure \ref{constant}.

\paragraph{Acknowledgements}
 The authors are indebted to B. Heller for fruitful comments and bibliographic suggestions.   
J.P.G. acknowledges the Centro Brasileiro de Pesquisas F\'isicas (CBPF) for financial support.

\end{document}